\documentstyle[12pt,aaspp4,astrobib,epsfig]{article}

\def\arcs{\ifmmode {'' }\else $'' $\fi}
\def\as{^{\prime\prime}}
\def\cf{{\it cf.\ }}
\def\etal{ et al.\ }

\def\cf{{\it cf.\ }}
\def\ie{{\it i.e.,\ }}

\def\erg{{\rm\thinspace erg}}
\def\km{{\rm\thinspace km}}

\def\kms{{\rm\,km\,s^{-1}}}

\def\Mpc{{\rm\thinspace Mpc}}
\def\Msun{\hbox{$\rm\thinspace M_{\odot}$}}
\def\s{{\rm\thinspace s}}

\def\ergps{\hbox{$\erg\s^{-1}\,$}}
\def\kmps{\hbox{$\km\s^{-1}\,$}}

\def\ltsima{$\; \buildrel < \over\sim \;$}
\def\simlt{\lower.5ex\hbox{\ltsima}}
\def\gtsima{$\;\buildrel > \over\sim \;$}
\def\simgt{\lower.5ex\hbox{\gtsima}}

\def\hmpc{$h^{-1}$\thinspace Mpc}
\def\msun{{\rm\,M_\odot}}

\begin{document}

\title{GALAXY EVOLUTION IN ABELL 2390}

\author{R. G. Abraham\altaffilmark{1}} 
\affil{Institute of Astronomy, University of Cambridge\\ 
Madingley Road, Cambridge CB3 OHA UK\altaffilmark{2}\\and\\
Dominion Astrophysical Observatory, Herzberg Institute of Astrophysics\\ 
National Research Council of Canada\\ 5071 W. Saanich Rd., Victoria B.C. 
V8X 4M6 Canada\\
abraham@ast.cam.ac.uk }

\author{Tammy A. Smecker-Hane\altaffilmark{1,3}, J. B. Hutchings}
\affil{Dominion Astrophysical Observatory, Herzberg Institute of
Astrophysics\\ National Research Council of Canada\\ 5071 W. Saanich
Rd., Victoria B.C. V8X 4M6 Canada\\ smecker@carina.ps.uci.edu,
hutchngs@dao.nrc.ca}

\author{R. G. Carlberg\altaffilmark{1}, H. K. C. Yee\altaffilmark{1}}
\affil{Department of Astronomy, University of Toronto,\\ Toronto,
Ontario, M5S 1A7 Canada\\ carlberg@moonray.astro.utoronto.ca,
hyee@makalu.astro.utoronto.ca}

\author{Erica Ellingson\altaffilmark{1}} \affil{CASA, University of
Colorado\\ Boulder, Colorado 80309-0389\\ e.elling@pisco.colorado.edu}

\author{Simon Morris\altaffilmark{1}, J. B. Oke, Michael
Rigler\altaffilmark{1}} \affil{Dominion Astrophysical Observatory,
Herzberg Institute of Astrophysics\\ National Research Council of
Canada\\ 5071 W. Saanich Rd., Victoria B.C. V8X 4M6 Canada\\
simon@dao.nrc.ca, oke@dao.nrc.ca, mrigler@microsoft.com}

\altaffiltext{1}{Guest Observer, Canada France Hawaii Telescope,
which is operated by the NRC of Canada, CNRS of France, and the
University of Hawaii.}

\altaffiltext{2}{Current address.}

\altaffiltext{3}{Current address: Department of Physics, University
of California, Irvine, CA 92717.}

\begin{abstract}

The galaxy population in the intermediate-redshift ($z=0.228$) rich
cluster Abell 2390 is investigated.  We present velocities, colors,
and morphological information for an exceptionally large sample of 323
galaxies (216 cluster members) in a 46$^\prime \times 7^\prime$ (6
$h^{-1}$ Mpc $\times$ 1 $h^{-1}$ Mpc) strip centered on the cD galaxy.
This sample of confirmed cluster members is second only to that for
the Coma cluster in terms of sample size and spatial coverage in the
cluster rest frame, and is the first to trace the transition between a
rich cluster and the field at intermediate redshift.  The galaxy
population in the cluster changes gradually from a very evolved,
early-type population in the inner 0.4 \hmpc\ of the cluster to a
progressively later-type population in the extensive outer envelope of
the cluster from 1 to 3 \hmpc\/ in radius.  Radial gradients in galaxy
$g-r$ color, 4000 \AA\/ break, H$\delta$ and [O II] line strengths and
morphology are seen in the cluster, and are investigated by comparing
the data to models computed with the GISSEL spectral synthesis
package.  The results suggest that the cluster has been gradually
built up by the infall of field galaxies over $\sim 8$ Gyr and that
star formation has been truncated in infalling galaxies during the
accretion process.  The morphological composition of the cluster is
shown to be consistent with such a scenario.  If true for other
clusters, infall-truncated star formation as seen in Abell 2390 may
explain both the Butcher-Oemler effect and the large fraction of S0
galaxies in clusters.  Only $\simlt$5\% of the galaxies observed in
Abell 2390 exhibit evidence for star formation at levels stronger than
those seen in typical late-type systems.  This suggests that
starbursts do not play a major role in driving cluster galaxy
evolution at the redshift of Abell 2390, although infall-induced
starbursts leading to truncated star-formation may have played a role
in the earier history of the cluster. Evidence is found for at least
one subcomponent on the West side of the cluster, which is likely to
be infalling at the epoch of observation.
\end{abstract}

\vfill\eject

\section{INTRODUCTION}

\noindent

Within the hierarchical gravitational instability theory, clusters of
galaxies are created by merging of smaller clusters. Such merged
clusters may have no memory of their initial conditions, or they may
retain the radial gradients of the progenitors.  More specifically,
``violent relaxation''~\cite{Lynden-Bell:1967} erases nearly all
memory of the initial structures (but see \citeNP{Quinn:1986}),
whereas steady accretion formation~\cite{Gunn:1972} builds the
cluster slowly and continuously over a Hubble time, so that the
orbits remain stratified in a radial age sequence.  Nearly equal mass
mergers will be quite ``violent'', but the steady accretion of
individual galaxies and small groups on to a pre-existing cluster
leaves the structure of the initial cluster relatively unaffected.
The ability of the cluster to continue to accrete material at low
redshift depends on the initial overdensity profile of the cluster
and $\Omega$~\cite{Gunn:1972}.

If a cluster is uniformly mixed then we presume that the last merger 
was quite violent.  Cluster galaxies are generally red, and the 
combination of their mean color and the {\em dispersion} in color 
allows a limit to be put on how closely in time the galaxies formed, 
and thus on the epoch of initial galaxy formation.  Observed gradients 
in galaxy populations can be used to trace the cluster's accretion 
history, and to test simple galaxy evolution models.  There is likely 
to be a close relationship between population gradients in clusters 
and the ``Butcher-Oemler Effect'' (the increase in the fraction of 
blue cluster members with redshift).  HST observations of the 
intermediate redshift clusters CL 0939+4713 ($z=0.41$; 
\shortciteNP{Dressler:1994}), Abell 370 ($z=0.39$; 
\shortciteNP{Couch:1994}), and AC 114 ($z=0.31$; 
\shortciteNP{Couch:1994}) have shown that in these three clusters the 
relative fraction of spirals, S0 galaxies, and ellipticals is similar 
to that seen in the {\em field} at the current epoch.  These 
observations suggest that the Butcher-Oemler effect is mostly due to 
an excess population of late-type systems, and not due to a population 
of starbursting, early-type systems.  If CL 0939+4713, Abell 370, and 
AC 114 (the only clusters for which morphological studies at $z \sim 
0.3 - 0.4$ have been published) are representative of rich clusters at 
intermediate redshift, then the cluster galaxy population has evolved 
since $z\sim 0.4$ into the early-type-dominated population seen in 
nearby clusters \cite{Dressler:1980}.  The narrow color-magnitude 
envelope for the red galaxies in the clusters studied with HST led 
both \shortciteN{Dressler:1994} and \shortciteN{Couch:1994} to 
conclude that the old population in these clusters is similar to 
ellipticals and S0s seen at the current epoch, and hence that galaxy 
evolution in clusters is dominated by the fading or destruction of 
cluster spirals.

Other evidence for cluster galaxy evolution was discovered by
\citeN{Dressler:1983}, who found that a substantial number of galaxies
in intermediate redshift clusters have enhanced Balmer absorption for
their color.  These authors coined the term ``E+A galaxy'' to describe
these objects, noting that their spectra could be matched with a
mixture of an elliptical galaxy and an A-type stellar population, and
that their colors were intermediate between those of early-type
systems and spirals.  Since Balmer lines are enhanced as the
main-sequence turnoff moves through A stars, which occurs $\sim 1$ Gyr
after star-formation (or a starburst) ceases, the spectroscopic
characteristics and colors of E+A galaxies have led most authors to
conclude that E+A systems are the remnants of starbursts that have
recently occurred in old galaxies.  However, others ({\it e.g.}
\citeNP{Couch:1987}) show that it is hard to distinguish between 1 Gyr
old starbursts in early-type systems and late-type systems whose
star-formation has simply been truncated without an initial starburst,
and in view of this ambiguity prefer the term ``H$\delta$-strong''
(HDS) to ``E+A''.  More recently, \citeN{Charlot:1994} have shown
that, solely on the basis of optical colors and spectra, it is
virtually impossible to distinguish between a major starburst in an
elliptical and one in a spiral if star formation ceases after the
burst.  The only unambiguous way to identify these starbursts is to
catch them {\em during} the burst when their [O~II] emission is strong
and their colors are blue.

In this paper we examine the galaxy population in the intermediate
redshift cluster Abell 2390 ($\alpha_{\rm 1950} = 21:51:14.3$,
$\delta_{\rm 1950} = +17:27:34.9$, $z=0.228$).  The data were obtained
as part of the Canadian Network for Observational Cosmology (CNOC)
dynamical survey of X-ray-luminous clusters of galaxies
\shortcite{Carlberg:1994}.  Abell 2390 is a large, rich cluster with a
sizable hot intracluster medium ($L_x=5.5 \times 10^{44}$ erg
s$^{-1}$; \citeNP{McMillan:1989}).  We use the observed colors,
spectral features and morphologies of the galaxies to put limits on
the star-formation histories of cluster galaxies and on the cluster's
accretion history. The plan of this paper is as follows. In \S2, we
describe our spectroscopic and photometric data and analysis, and
outline the automated procedure used to obtain the morphological
classifications. In \S3, we discuss the rich spatial and velocity
structure of the cluster. We demonstrate that Abell 2390 contains a
very old, red, centrally condensed component, and that galaxies in the
extensive outer envelope of the cluster show a radial gradient in
which bluer, later type galaxies are found systematically at larger
radii.  We also identify a distinct group of galaxies (red as well as
blue: the `NW Group') that appear to constitute a small cluster that
is merging with Abell 2390. In \S4, we discuss the galaxies with
current star formation ([O II] emission line galaxies) and those
galaxies which have recently experienced a significant decrease in
their star formation rates (strong Balmer absorbers).  In \S5, we
discuss how we have used the GISSEL spectral synthesis package
\shortcite{Bruzual:1993} to calculate the line strengths and colors of
galaxies for various galaxy evolution models.  In \S6, we compare the
data with these models.  In \S7, we show that the gradients in galaxy
color, spectroscopic line measures, and morphology are consistent with
a scenario in which the the outer part of the cluster is built up
slowly over a Hubble time, while star formation is truncated in
infalling galaxies. In this picture, the radial gradients in Abell
2390 are due to systematic changes in the ages of the stellar
populations in the galaxies as a function of radius.  Abell 2390 is
similar to other rich clusters at intermediate redshifts, and we
speculate that truncated star formation may play an important role in
the Butcher-Oemler effect and lead to the formation of cluster S0
galaxies. We also discuss evidence for merging and interactions in our
imaging data.  Our conclusions are summarized in \S8.

\section{OBSERVATIONS AND DATA REDUCTION}

\subsection{Optical spectroscopy and imaging}

Abell~2390 was observed in June and October 1993 using the MOS arm of
the MOS/SIS multi-object spectrograph~\cite{LeFevre:1993}, mounted at
the f/8 focus of the 3.6 meter Canada-France-Hawaii Telescope (CFHT).
The observing strategy and spectroscopic reduction process are
described in detail in \citeN{Yee:1996a}. Images in Cousins R and Gunn
$g$ filters were also obtained through the spectrograph in order to
design the aperture masks; these are described in Section 2.3
below. Photometry was calibrated to the Gunn $g$ and $r$ system.  The
extracted spectra cover the wavelength range 4300 -- 5600~\AA\/ (3500
-- 4550~\AA\/ in the rest frame of the cluster) with a dispersion of
3.45~\AA\//pixel and a resolution of 15.5~\AA\/.  The spatial scale at
the rest frame of the cluster is $430\as /$(\hmpc).  A total of 323
redshifts covering a 46$^\prime \times 7^\prime$ strip (five slightly
overlapping MOS fields centered on the cD galaxy) were obtained.  Of
these, 233 galaxies have redshifts in the range $0.20 \leq z \leq
0.265$, and 216 of these are considered to be cluster members (our
criteria for determining cluster membership are described in \S3.1).
A complete catalog of cluster and field galaxies in our sample is
given in \citeN{Yee:1996b}.  Table 1 lists photometry, velocities,
identification numbers (from \shortciteNP{Yee:1996b}), and new
spectroscopic and morphology measurements for the galaxies discussed
in the present paper.

As with all surveys, selection effects play an important role in
defining the present sample.  In order to minimize selection anti-bias
against close pairs, spectra for the crowded inner fields were
obtained with 2 or 3 aperture masks.  Even so, selection biases of
order 30\% remain over the surface of the cluster, but these can be
accurately calibrated from our known selection function.  A detailed
completeness plot for the current dataset is given in
\citeN{Yee:1996b}.  As a rough guide, the cumulative completeness of
the spectroscopic sample is over 80\% to $r = 20.0$ mag and $\sim$60\%
to $r = 21.0$ mag.  (The corresponding differential completeness rates
at these magnitude limits are 74\% and 30\%, respectively).  In
\citeN{Yee:1996b} it is shown that the present sample is unbiased with
respect to color for $r<21$ mag.  Beyond $r = 21$ mag completeness
drops rapidly and is a strong function of the color, because redshifts
are more easily derived from blue galaxies with [O II] emission lines.
Hence in this paper the sample will be restricted to galaxies with
$r\leq21.0$.  Since $r=21$ mag corresponds to $M_V\sim -19$ mag at
$z=0.23$ (assuming $H_{o}=75 \, \kms \, {\rm Mpc}^{-1}$), the
spectroscopic dataset probes $\sim 1.5-2$ mag fainter than $M^\star$
in the cluster luminosity function (assuming $M^\star_{B_T}=-20.3$
mag, as found by \citeNP{Efstathiou:1988}).

The completeness limits for object detection on our images varies from
field to field, as a result of differing integration times on our CCD
frames (900s in the central field, 600s in the outer fields).  The
detection completeness limit spans a range from 23.5 to 24 mag for
$r$, and 23.2 to 24 mag for $g$.  A variable PSF across the MOS field
limits successful star-galaxy separation to objects with r $\le$ 23.5
mag.  (The quality of the images was limited by the MOS optics rather
than by observing conditions; the FWHM of the PSF varied between
$0.9\as - 1.3\as$).  Taken together, the imaging data are complete to
$M_U\approx-17.0$ mag and $M_V\approx-16.6$ mag in the rest frame of
the cluster.

\subsection{Spectroscopic Measures}

The two-dimensional spectral images were cleaned of cosmic rays and
extracted (with variance-weighting) using the IRAF ``apextract''
package.  Spectra were wavelength-corrected, flux-calibrated and
extinction-corrected to bring all spectra to the same system.
Redshifts and the confidence parameter $R_{xcor}$ were derived using
cross-correlation techniques.  Details of the data reduction procedure
are given in \citeN{Yee:1996a}, and the final
photometric/spectroscopic catalog, together with completeness and
selection functions, is presented in \citeN{Yee:1996b}.  In this paper
additional measurements will be presented for all spectra in
\citeN{Yee:1996b} with derived redshifts.  These measurements include
the 4000\AA\/ break (D4000, \cf \citeNP{Hamilton:1985}), and the
equivalent widths of [O II] 3727\AA~and H$\delta$.  In the present
paper the signal-to-noise ratio (S/N) of the spectra are quantified in
terms of the mean S/N in the rest frame 4050 -- 4250 \AA\/ region.

Equivalent widths at the observed redshift of each spectrum were
calculated assuming fixed rest wavelengths for each line and pair of
continuum regions.  The observed equivalent widths were then converted
to the rest frame by dividing by ($1+z$).  The definitions of the line
and continuum regions used in the present work are shown in Table 2.
The definitions are based on those of \citeN{Dressler:1987} but were
modified to be compatible with the spectral resolution of both the
observed data and the model spectra generated by the GISSEL spectral
synthesis package \shortcite{Bruzual:1993}.  The equivalent widths for
the observed and model spectra in this paper were computed in
identical ways.  The noise in the spectrum of the sky background was
used in calculating the measurement error for each spectral feature
following the formulae in \shortciteN{Bohlin:1983} Appendix A.
Internal checks based on multiple observations of the same galaxies
through different spectroscopic masks verify that the errors
calculated in this way are a good description of the uncertainties in
our spectroscopic measurements.

One of the most important line measures is H$\delta$, because it is
used to identify galaxies which have recently ceased vigorous star
formation.  The actual width of the line varies as the stellar
population evolves and metallic absorption lines begin to appear in
the nearby continuum.  Extreme care has been taken in measuring
H$\delta$ in both the observed spectra and the models, and our methods
have been tested against high resolution extant spectra of stars and
galaxies.  The continuum at H$\delta$ was fitted using relatively wide
bandpasses and then re-fitted rejecting the points lower/higher than
-5/+13 times the average deviation from the initial fit.  This simple
iterative procedure nicely masks out absorption features and noise
spikes and leads to a more robust definition of the true continuum.
In addition, equivalent widths for two line bandpasses that are 28
\AA\/ and 40 \AA\/ wide and centered on H$\delta$ were calculated
(H$\delta_{\rm nar}$ and H$\delta_{\rm wide}$, respectively).  For the
H$\delta$ measurement, the narrower bandpass is adequate for
absorption $< 2$ \AA, but when the absorption is stronger and the
wings of the line broader then the equivalent width calculated with
the wider bandpass needs to be used in order to adequately span the
full range of H$\delta$ absorption strengths. Thus the definition for
H$\delta$ equivalent width used in this paper is: H$\delta$ =
H$\delta_{\rm nar}$ if H$\delta_{\rm nar} < 2$ \AA, and H$\delta$ =
H$\delta_{\rm wide}$ if H$\delta_{\rm nar} > 2$ \AA.

\subsection{Morphological parameters}

The morphologies of the galaxies in Abell 2390 were determined using
the automated system of \citeN{Abraham:1994}, which is quite similar
to the method proposed by \citeN{Doi:1993}.  The fundamental
parameter of this system is a galaxy's central concentration of
light, $C$.  This parameter is determined from the intensity-weighted
moments of the galaxy images.  It is emphasized that $C$ values are
morphological {\em measurements}, rather than classifications.
Because it is based on the central concentration of light, the
\shortciteN{Abraham:1994} system bears more similarity to Morgan's
Yerkes system \cite{Morgan:1958} than it does to the standard Hubble
system. However, it has long been known \cite{Shapley:1927} that
central concentration measurements can be used to distinguish
effectively between early and late galaxy types. Determining more
subtle distinctions between galaxies on the Hubble sequence (eg.
distinguishing ellipticals from S0 galaxies) cannot be done from $C$,
but in any case such distinctions cannot be made visually at the
redshift of Abell 2390 from our images.  The main advantages of the
automated classification system are (a) objectivity, (b)
applicability at redshifts where classifications based on the Hubble
system are difficult to make, and (c) the existence of well-defined
uncertainties on the morphological measurements.  As described in
\shortciteN{Abraham:1994}, the values of $C$ determined for each
galaxy result from a two-step reduction process. Direct measurements
of $C$ were first made using code added to the FOCAS galaxy detection
package~\cite{Valdes:1982}.  In the second stage of the reduction
process these $C$ measurements were corrected for the effects of
seeing degradation, using the results from Monte Carlo simulations.

For un-blended and fairly smooth galaxies, the central concentration 
of light was determined within an area enclosed by an isophotal 
surface brightness limit that is $2.5 \sigma$ above the noise in the 
sky background.  Since our images span a range of exposure times 
between 600s and 900s, the surface brightness of the limiting 
isophotes for the unblended galaxy population range between 
$\mu_r=24.1-24.5$ mag/arcsec$^2$.  Simulations indicate that a change 
of $\sim 0.4$ mag/arcsec$^2$ in the isophotal limit can change the 
measured $C$ values by $\sim 0.03$, which is small compared to the 
measurement errors inherent in determining $C$.  For blended or highly 
distorted galaxies, the limiting isophote determined by FOCAS is 
substantially higher than the $2.5 \sigma$ sky noise limit.  (FOCAS 
continuously raises the limiting isophotal level when a detected 
object's isophote encloses several maxima, until only a single maximum 
is enclosed.) Since galaxies with high limiting isophotes correspond 
to blended or distorted objects, high isophotal limits flag possible 
merger candidates.  Approximately 20\% of the cluster members had high 
limiting isophotes ($\mu_r < 23.5$mag/arcsec$^2$).  However, the 
majority of these objects correspond to smooth-looking galaxies with 
stars or faint galaxies superposed on the galaxy images, rather than 
to obviously distorted interacting or merging systems.  As discussed 
in Section 7 below, higher resolution images are required in order to 
determine unambiguously the rate of mergers and interactions in Abell 
2390.

\section{SPATIAL-VELOCITY STRUCTURE AND SUBCLUSTERING}

\begin{figure}
\centerline{\psfig{figure=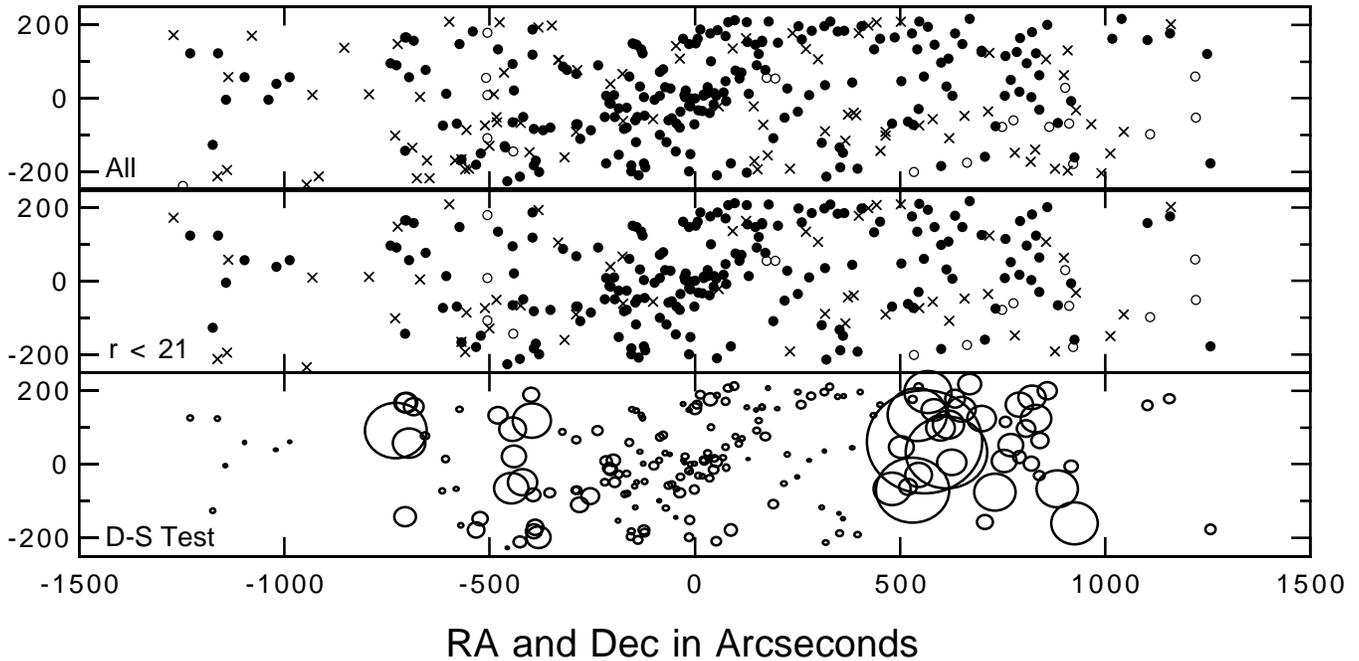}}
\caption{ The spatial distribution of galaxies in
  the Abell 2390 sample.  Solid circles denote cluster members, open
  circles denote ``near field'' galaxies (see text), and crosses
  denote field objects.  (Top) All galaxies with known redshifts.  (b)
  Galaxies brighter than $r=21$ mag (the restricted sample analyzed in
  the present work).  (c) The results from a Dressler-Shectman test
  for spatial-velocity substructure.  The radii of the circles are
  proportional to $e^{\delta}$, where $\delta$ is the
  Dressler-Shectman subclustering estimator (Dressler \& Shectman
  1988).}
\label{fig:spatial} 
\end{figure}

\begin{figure} 
\centerline{\psfig{figure=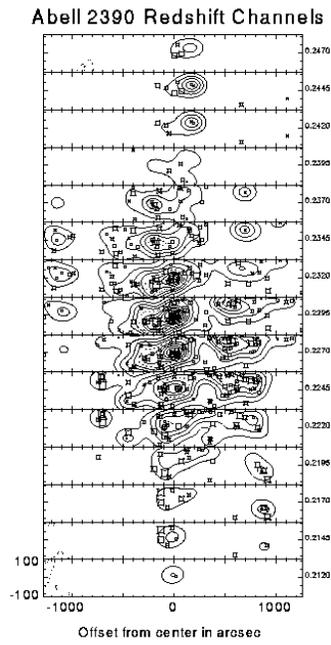,width=3.4in}}
    \caption{ Redshift channels showing the
    distribution of cluster members in both spatial and velocity
    dimensions. The rich structure of our dataset in spatial-velocity
    space is obvious.  The distribution of points in each panel has
    been smoothed with an angular Gaussian of FHWM 80\arcsec~in the
    spatial dimensions and 0.0025 in the redshift dimension. The area
    of each plot symbol is proportional to the luminosity of the
    galaxy.}
\label{fig:channel}
\end{figure}

The spatial distribution of the spectroscopic sample is shown in 
Figure~\ref{fig:spatial}.  The central core of the cluster spans 
approximately 400\arcs\ (0.93 \hmpc) on either side of the cD galaxy 
(which is assumed to mark the spatial origin of the cluster 
throughout this paper), before breaking up into structures whose 
proximity in velocity space (discussed below) suggests that they are 
gravitationally bound to the cluster.  These structures are 
illustrated more clearly in Figure~\ref{fig:channel}, which shows the 
spatial map of the cluster sliced into redshift ``channels''.

\begin{figure} 
\centerline{\psfig{figure=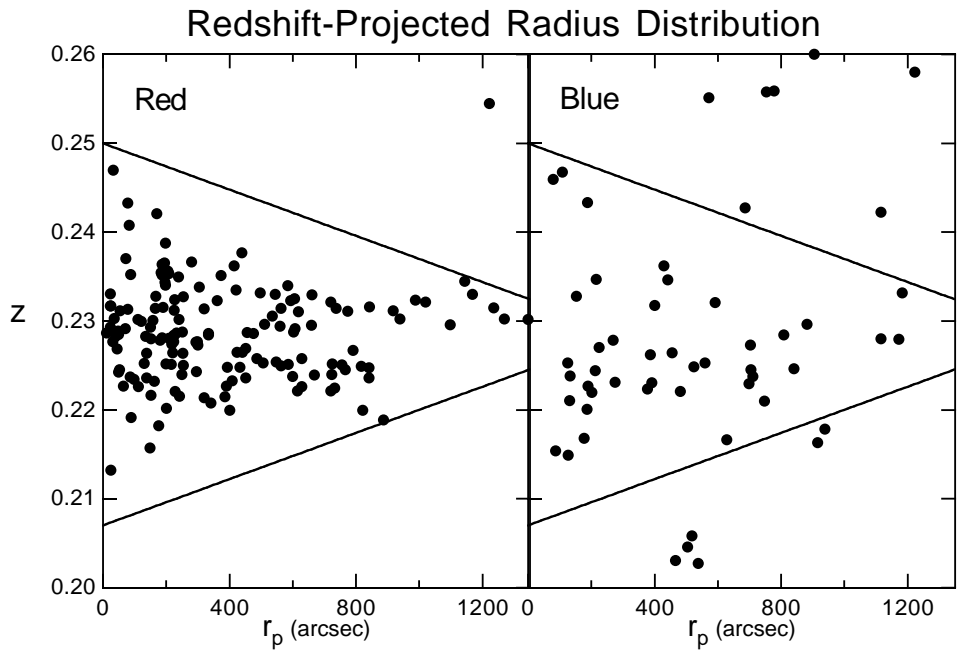,width=5in}}
  \caption{ The redshift-projected radius distribution for
  red (left) and blue (right) galaxies.  The well-defined distribution
  of red cluster members provides the basis for our definition of
  cluster and field membership (solid lines).  See text for our
  definition of red and blue objects.}\label{fig:cusps}
\end{figure}

For this paper, cluster membership is defined in a simple
(statistical) way using the redshift--projected radius distribution of
red galaxies shown in Figure~\ref{fig:cusps} (the precise definition
for ``red'' and ``blue'' galaxies adopted in the present work is given
in the next section).  Red galaxies appear well-separated from the
field and are assumed to trace the boundaries of the cluster. The two
straight lines in Figure~\ref{fig:cusps} are used to define the border
between cluster and field.  This definition is purely empirical, but
the clear absence of red galaxies outside the boundaries suggests that
a cut made solely on the basis of redshift would include a further
$\sim$15 high velocity blue galaxies that are unlikely to know about
the cluster structure, even for high $\Omega$ \cite{Regos:1989}.
These blue objects may be infalling, and hence the field galaxy
population is subdivided into ``far field'' objects that are outside
the $0.2 \leq z \leq 0.265$ redshift range, and ``near field'' objects
that are inside the redshift range but outside the boundaries shown in
Figure~\ref{fig:cusps}.  The adopted definition for cluster members
may still include a few infalling galaxies in the cluster, but the
main goal of this paper is to highlight the transition in galaxy
properties from the cluster center to the outer cluster while
minimizing contamination from the field at large radii.  The degree of
contamination by field galaxies due to the ``finger of god'' effect
(the elongation in redshift space of clusters due to their high
velocity dispersion) can be estimated as follows \cite{Koo:1988}. The loci
in Figure~3 trace out a total volume of $\sim 500h^{-3}$ Mpc$^3$ over
the rectangular configuration of our dataset, which is considerably
larger than the physical volume occupied by the cluster.  The degree
of contamination by field galaxies is dependent mostly upon the
richness and geometrical configuration of the large scale structure in
which the cluster is embedded. If one assumes a galaxy density of $<1$
galaxy/100 Mpc$^3$ (typical of the field) then the expected
contamination is only a few galaxies, which is negligible given our
large sample. Of course it is possible that the cluster is embedded in
a particularly rich component of large scale structure (which would
result in greater contamination), but this seems unlikely from the
redshift ``pie'' diagram given in Carlberg et al. (1995).  It is
emphasized that the conclusions in this paper are not strongly
sensitive to the precise shape of the boundaries in
Figure~\ref{fig:cusps}. Our $r<21 $mag sample is therefore composed of
199 cluster members, 14 near field galaxies, and 56 field galaxies.

The previous figures suggest that the cluster is composed of two or
three main clumps, embedded in a sparsely-distributed envelope of
galaxies.  This visual impression is supported by the results from a
Dressler-Shectman test (the bottom panel in Figure~\ref{fig:spatial}),
which indicates that global substructure is significant with over
99.9\% confidence. The strongest correlated motions occur over an
extended region on the West side of the cluster. The Dressler-Shectman
test is not sensitive to structure that is resolved in velocity space
but is spatially superposed, so the test is not ideal for
investigating velocity correlations amongst galaxies in the inner part
of Abell 2390.  For example, there is a remarkably low velocity spread
among the innermost group of red galaxies (see Figure~\ref{fig:cusps})
in the cluster, suggesting that the inner part of the central
component of the cluster is dynamically ``cold''.  This is not evident
in the results from the Dressler-Shectman test.

\begin{figure}  
\centerline{\psfig{figure=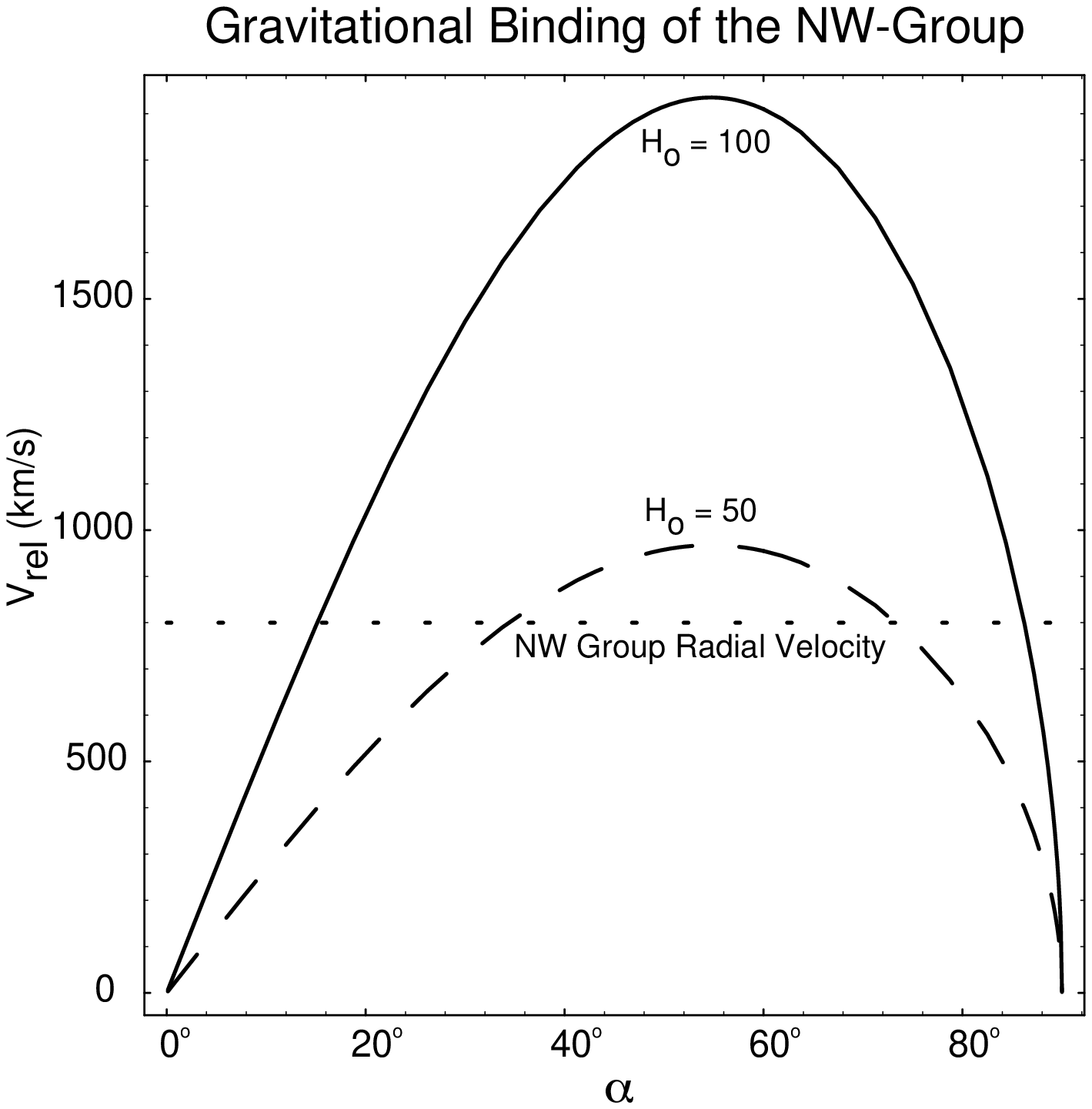,width=5in}}
\caption{ The bound and unbound regions in the
    $(V_{rel},\alpha)$ plane for the ``NW-Group'' in Abell 2390.
    $\alpha$ is the angle between the plane of the sky and the line
    joining the centers of the two components. All points below the
    curves are gravitationally bound. The calculations assume a mass
    for the cluster of $1.1 \times 10^{15}\Msun$, the lower limit to
    virial mass of the cluster determined by Carlberg et al. (1995). If
    the Hubble constant is high or if dark matter does not trace the
    luminous matter in Abell 2390 (as claimed by Carlberg et al. 1995)
    the NW-Group is likely to be bound.} \label{fig:bound}
\end{figure}

While the general presence of substructure in Abell 2390 seems clear,
the identification of small physical subclumps from cluster catalogs
is difficult to undertake reliably \cite{West:1990}, and in this paper
we will concentrate only on the two most obvious components seen in
Figures~\ref{fig:spatial} and \ref{fig:channel}.  The first of these
is the main cluster component centered on the cD.  This dominates the
central 400\arcs~of the cluster, and in the next section it is shown
that the red colors, evolved spectra, and early-type morphologies in
this component are very uniform.  The second obvious component in our
spatial-velocity diagram is the ``NW Group'', the (lower redshift)
concentration to the northwest of the main central component of the
cluster, at projected radius $\simeq 650\as$.  There are some 20
galaxies in this group, whose mean color is also significantly redder
than that of their surroundings, and whose co-added spectrum shows
them to be similar to the evolved galaxies at the cluster center. It
appears likely that this is the core of a smaller cluster that is
being accreted on to the main component of Abell 2390. Indeed, a
simple two-body analysis suggests that this group is at least weakly
bound to the cluster.  In Figure~\ref{fig:bound} we show the result of
a two-body calculation based upon the simple criterion derived by
\citeN{Beers:1982} for gravitational binding in subclumps:
\begin{equation}
\label{eqn:bound} V^2_{rel} R_p < 2 G M_{tot} \sin^2\alpha
\cos\alpha, \end{equation}

\noindent where $M_{tot}$ is the total cluster mass, $V_{rel}$ is the
relative velocity between the two components along the line of sight,
$R_p$ is their projected separation, and $\alpha$ is the angle between
the plane of the sky and the line joining the centers of the two
components. The NW Group is bound for most choices of $H_{\rm o}$ and
$\alpha$.

\section{THE GALAXY POPULATION}

\subsection{Blue Galaxies}

\begin{figure}
\centerline{\psfig{figure=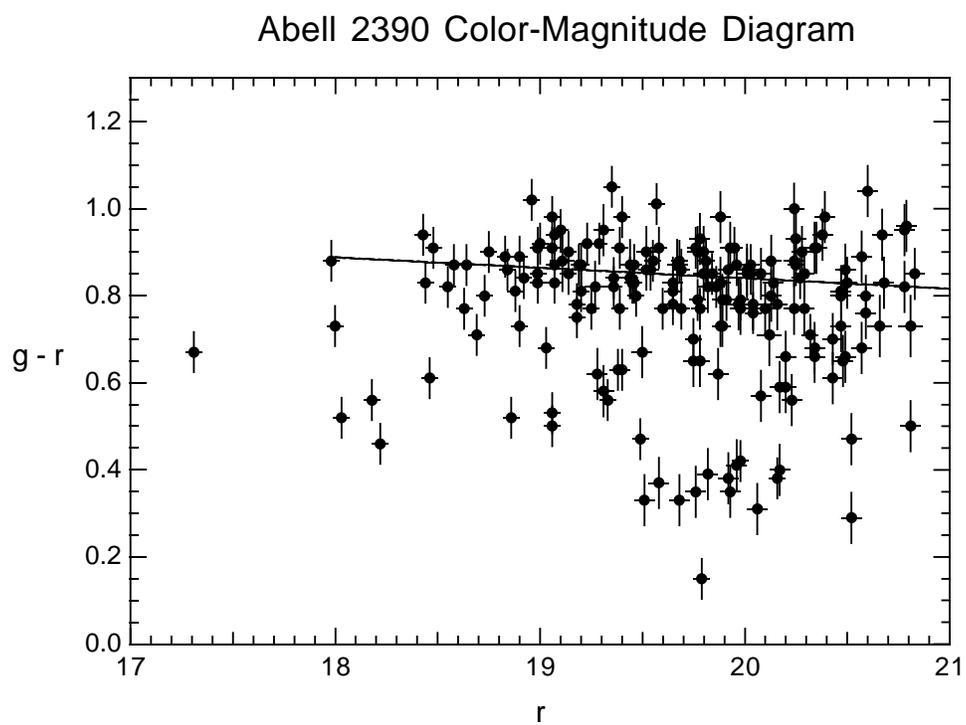,width=5in}}
 \caption{ Color-magnitude diagram for cluster galaxies.  
    The line is the best fit to the E/S0 color-magnitude relationship
    of cluster members.  See text for details.}\label{fig-cmd}
\end{figure}

The color-magnitude relationship in Abell 2390 is shown in 
Figure~\ref{fig-cmd}.  Iteratively fitting the E/S0 sequence as a 
straight line using a chi-squared minimization and excluding cluster 
galaxies $\pm 0.15$ mag outside the fit gives the following 
color-magnitude relationship:

\[(g-r) = 1.32 (\pm 0.14) - 0.024 (\pm 0.007) \times r.\] 

\begin{figure} 
\centerline{\psfig{figure=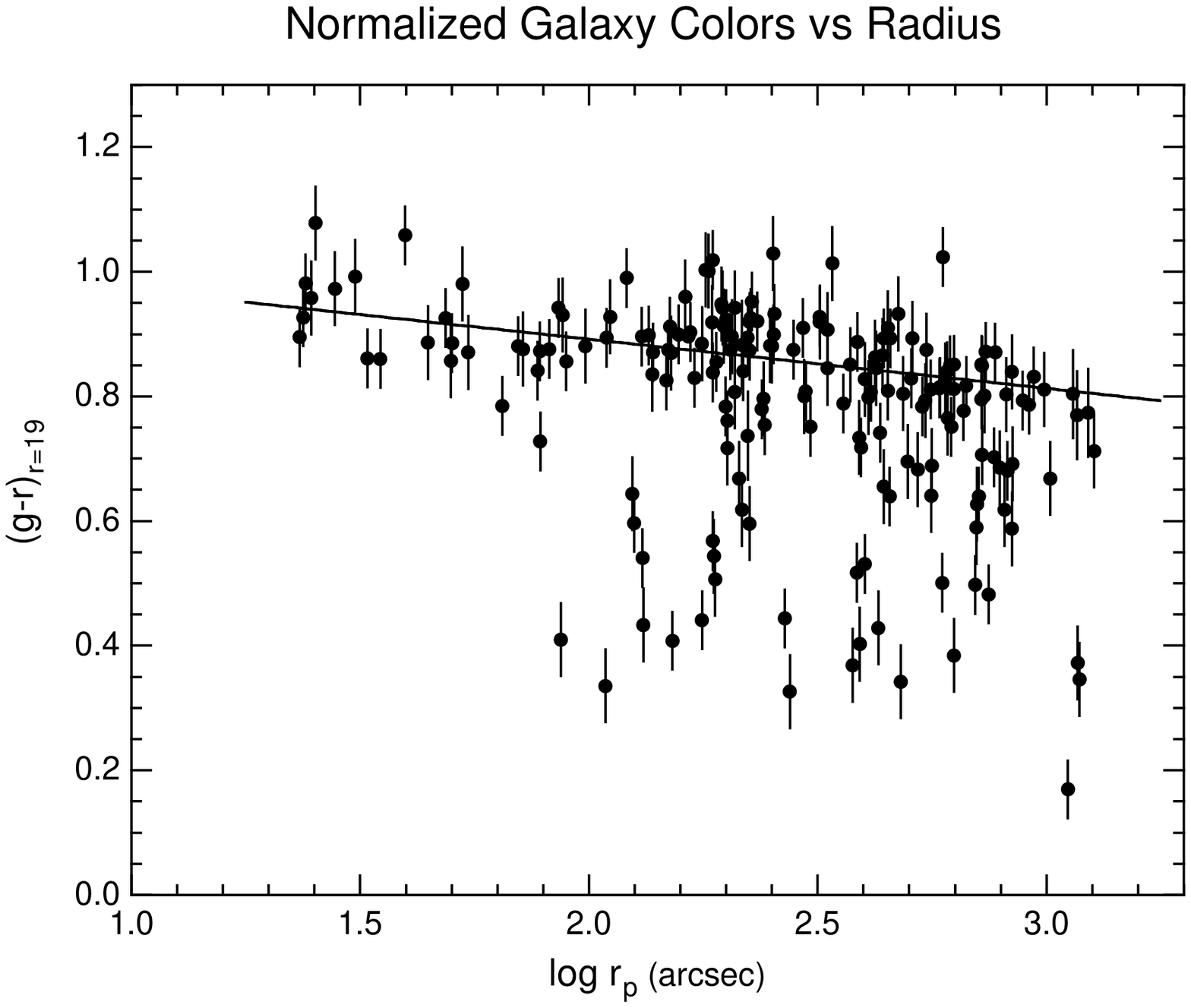,width=5in}}
\caption{ The color-projected radius relationship for
    cluster galaxies where color has been normalized to $r=19$ using
    the color-magnitude relationship.  The line is the best fit
    straight line to the red galaxy locus. } \label{fig-grrp}
\end{figure}

The scatter in this relationship is largely due to a strong radial 
color {\em gradient} in the color of the old galaxy locus in the 
cluster (Figure~\ref{fig-grrp}).  Normalizing colors to $r=19$ using 
the above color-magnitude relationship and computing the best fit line 
for the E/S0 ridge in the color-radius plane, we obtain the following 
relationship for the normalized color of the red galaxy population
as a function of projected radius \(r_{p}\):

 \[ (g-r)_{r=19} = 1.05 - 0.079 \, \log r_p \label{radeqn}\]

The color-radius relationship for the red galaxies is only 
approximately linear in $\log\, r_p$.  The slope of the relationship 
appears shallower for the inner cluster and steeper for $r_p \simgt 
200\as = 0.46$ \hmpc.  The spread in red galaxy colors is 
approximately 0.15 mag over the full range of radius.

The radial gradient in the colors of the red galaxies leads to the
following definition for the blue population.  {\em Blue objects are
defined to be those galaxies at least 0.25 mag bluer in $g-r$ than the
color of the red galaxy locus at the projected radius of each
individual galaxy}.  The cutoff at $\Delta(g-r)=0.25$ was chosen
because this corresponds closely\footnote{$\Delta(B-V)=0$ corresponds
to $\Delta(g-r)$=0.26 according to Equation \ref{eqn:bvgr}, which is
based on GISSEL models, while \citeN{Yee:1996b} give
$\Delta(g-r)$=0.23 based on the spectral energy distributions of
\citeN{Coleman:1980}.}  to the rest-frame $B-V=0.20$ offset adopted by
Butcher \& Oemler (1984).

\begin{figure} 
\centerline{\psfig{figure=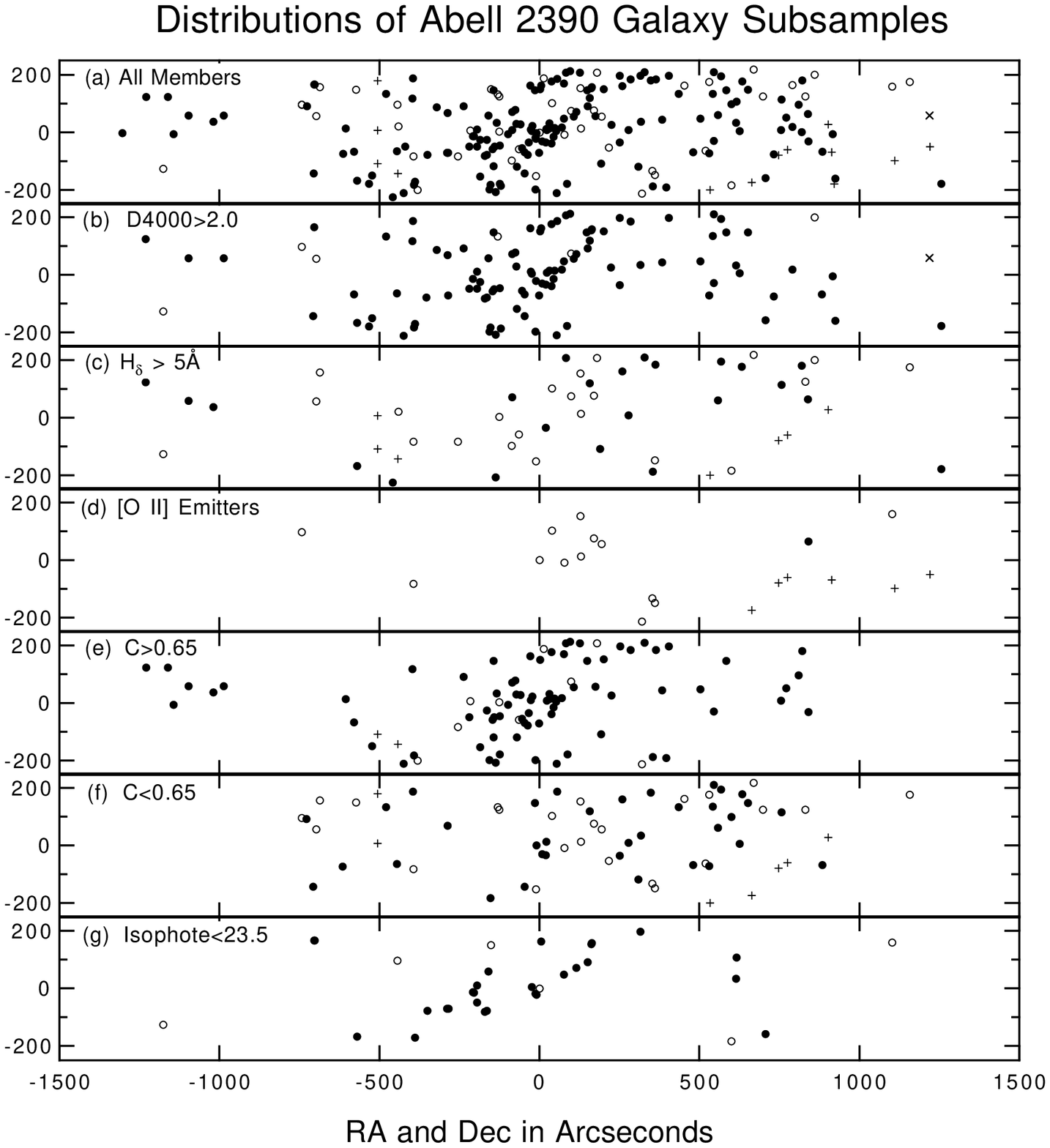,width=5in}}
\caption{ The spatial distribution of galaxy subsamples in
    Abell 2390.  Solid circles denote red cluster members, and open
    circles denote blue cluster members.  Red and blue near-field
    galaxies are denoted by crosses and plus signs, respectively.
    From top to bottom the panels show: (a) All cluster members and
    near-field galaxies.  (b) Objects with high 4000\AA~breaks.  (c)
    Galaxies with high H$\delta$ absorption.  (d) Emission line
    objects.  (e) High central concentration galaxies ($C>0.65$).
    These galaxies are likely to be early type.  (f) Low central
    concentration galaxies ($C<0.65$).  These are likely to be late
    type.  (g) Galaxies with limiting isophotes brighter than 24
    mag/arcsec$^2$, as determined by FOCAS.  These galaxies have
    nearby companions or morphological distortions preventing the
    clean separation of image components.}
\label{fig-spatial2} \end{figure}

The spatial distributions of red and blue objects are shown in panel
(a) of Figure~\ref{fig-spatial2}.  The distribution of colors in the
galaxy population is qualitatively similar to that seen in local
relaxed clusters, with a strong concentration of red galaxies in the
cluster center and an increasing fraction of blue galaxies as a
function of radius.  The NW Group is also dominated by red galaxies,
the reddest colors of which appear to be slightly ($\sim 0.1$ mag)
bluer than the reddest galaxies at the center of the main body of the
cluster.

\begin{figure} 

\caption{ Montage of spectra from different types and
    subsets of galaxies in A2390. Each panel is the sum of 8 -- 10
    representative spectra of S/N$>$6. The principal line features are
    identified and all wavelengths correspond to z=0.228.  The left
    side shows evolved populations, including the NW Group (from top
    to bottom, parent groups are D4000 $>2.2$; central 19 galaxies; NW
    group of 23). The right side shows younger populations, selected
    by H$\delta$ absorption ($>$4\AA) and color ($g-r<0.67$,
    H$\delta<2$).  Also shown are summed field galaxies
    ($0.27<z<0.33$), which have similar young population and blue
    color.  The field galaxy spectrum is truncated at red wavelengths
    because of the shifting required to bring field objects into the
    cluster rest frame.} \label{spectra}
\end{figure}

\subsection{Line Emitting Galaxies}

The distribution of [O~II] emitters (2$\sigma$ detections confirmed by
visual inspection) is shown in panel (d) of Figure~\ref{fig-spatial2}.
Fourteen cluster members, six near field galaxies, and fourteen field
galaxies exhibit [O~II] emission.  Assuming Poisson statistics, the
fraction of detected line-emitters is therefore estimated at only
$7\pm1\%$ amongst cluster members, $43\pm16\%$ in the near field
population, and $25\pm7\%$ in the field.  We note that the emission
line field sample may be biased with redshift due to the presence of a
band-limiting filter and because of the low quantum efficiency of the
CCD at blue wavelengths.  However, our detection of [O~II] in cluster
galaxies is not biased with redshift: we find emission in similar
numbers of cluster members with redshifts below and above $z=0.228$
(six and eight objects, respectively).

The spatial distribution of [O~II] emitters in the cluster is
remarkably non-uniform. Eleven cluster members and six near-field
objects are on the West side of the cluster, versus two cluster
members and no near-field objects on the East side.  This 17:2
West/East discrepancy in the positions of cluster and near field
line-emitters is not due to uneven sampling on either side of the
cluster.  In the field, eight [O~II] emitters are on the West side of
the cluster, versus 6 [O~II] emitters on the East side. The
distribution of non-[O~II]-emitting galaxies is also fairly uniform:
102 members, ten near field objects, and 29 field galaxies line on the
West side of the cluster, versus 95 members, four near-field objects,
and 26 field galaxies on the East side.

More meaningful counts can be determined by weighting objects by the
geometric sampling factor $W_{xy}$, given in \shortciteN{Yee:1996b},
which is simply the inverse of the geometric selection function at the
position of each galaxy.  This factor accounts for biases in slit
placement due to local crowding in the slit positions, as well as for
differences in the relative number of masks used to obtain spectra
across the face of the cluster (the easternmost field had only one
mask, whereas the westernmost field had two).  Incorporating these
geometric weights does not result in a significant change in the
West/East bias amongst line-emitters: 17.71 cluster and near-field
galaxies lie on the West side of the cluster, versus 1.39 galaxies on
the East side.  The changes also remain small in the control sample of
field [O~II] emitters (6.84 objects lie to the West of the cluster
center, versus 7.66 objects to the East) and non-[O~II] emitters
(109.42/10.06/29.94 to the West of the cluster center, versus
94.85/2.97/26.64 objects to the East, amongst members/near-field/field
galaxies).

\subsection{H$\delta$-Strong Galaxies}

H$\delta$ absorption with an equivalent width greater than 5\AA~was
detected in 45 galaxies (23\% of the cluster sample).  The
distribution of these objects is shown in panel (c) of
Figure~\ref{fig-spatial2}, and is centrally concentrated, although
objects with H$\delta > 5$\AA~are scarce near the very center of the
cluster ($r_{p}<100\as$).  Twenty-two of these objects are blue and
are likely to be late-type systems.  The remaining 23 objects (12\% of
the cluster sample) have abnormally strong H$\delta$ for their colors,
and are classified as ``HDS systems''\footnote{Note that the taxonomy
used to denote these objects can be somewhat confusing, since blue
objects with H$\delta > 5$\AA~in the present sample certainly have
strong H$\delta$ absorption.  However, the H$\delta$ absorption in
these objects is not {\em abnormally\/} strong for their color, and
hence these galaxies are not classified as HDS systems.}.

The H$\delta>5$\AA~criterion used in the definitions of the HDS sample
has been chosen rather conservatively because of the comparatively low
signal-to-noise level of the spectra in the present sample.  In
Section 6.2 it is shown that models for H$\delta$ evolution predict
$H\delta \sim 1{\rm \AA}$ for highly evolved early-type systems, and
that $\sim 35$ evolved objects have $2{\rm \AA}<{\rm H}\delta<5$\AA.
Many of these objects are likely to be HDS systems.  However,
measurements of $H\delta$ in the present sample have typical
uncertainties of $\sim 2$\AA, and reliable classifications of {\em
individual\/} objects in the $2{\rm \AA}<{\rm H}\delta<5$\AA~regime
cannot be made (although statistical comparisons between the
distribution of these galaxies and model predictions are meaningful).
A fairly crude estimate of the number of HDS systems that are being
missed because of low signal-to-noise can be obtained by co-adding
subsamples of low signal-to-noise spectra.  These spectra were first
shifted to the same arbitrary wavelength ($0.2280$) before being
summed (Figure~\ref{spectra}).  The summed spectra shown in this
figure were generally composed of 5 to 10 spectra of S/N$>6$,
individually inspected for flaws, and of approximately equal signal.
The co-added spectra suggest that a significant fraction (around half)
of the red galaxies in the $2{\rm \AA}<{\rm H}\delta<5$\AA~regime are
HDS systems.  However, co-adding spectra is a procedure subject to
many possible systematic errors, and this exercise should only be
regarded as providing a rough upper limit of $\sim 20{\rm \%}$ to the
fraction of HDS systems in the cluster.  Therefore the HDS fraction in
Abell 2390 is estimated to be 12-20\%.

\section{GALAXY POPULATION MODELS}

In order to investigate the nature of the galaxy population in Abell 
2390, GISSEL was used to model the evolution of galaxy spectra and 
colors.  A \citeN{Scalo:1986} initial mass function (IMF) was assumed, 
with a stellar mass range from $0.1 \msun$ to $65 \msun$.  The results 
vary only slightly with different IMF choices, except if the IMF mass 
range is severely altered.  (A starburst forming only massive stars is 
shown below as an example.) Colors and D4000 in the observed frame, 
and equivalent widths in the rest frame, were computed in the same 
manner as for the observations.  The free parameters in our analysis 
were the star-formation rates and ages of the galaxies, as well as the 
reddening in the rest frame of the cluster.  A foreground Galactic 
reddening in the direction of Abell 2390 of $E(B-V) = 0.075 \pm 0.005$
\cite{Burstein:1982} was assumed, which corresponds to a foreground reddening of
$E(g-r) = 0.084$ using the calibration of \citeN{Kent:1985}.  The 
possible effects of heavy dust obscuration localized to vigorous 
star-forming regions (as in IRAS galaxies) have not been modeled.  
Also, because the GISSEL models are based upon a library of solar 
metallicity spectra, it is not possible to explore the effects of 
metallicity.  (The effects of metallicity are currently being 
investigated with Bressan et al.  [1994] \nocite{Bressan:1994} 
models.) However, the models of
\shortciteN{Worthey:1994} have been used to quantify the dependence of 
color and D4000 on metallicity for elliptical galaxies.  In addition, 
the GISSEL models do not include an HII region component.  In 
star-forming galaxies, Balmer emission from HII regions will increase, 
and may overwhelm, the stellar Balmer absorption.  Therefore, the 
calculated Balmer absorption equivalent widths for star-forming 
galaxies should be considered as upper-limits when comparing with real 
galaxies.  The effect of including emission lines on the integrated 
colors is negligible.

The default definitions of the Gunn $g$ and $r$ filters in the GISSEL 
filter list have been used.  These are based on the total transmission 
of the Palomar 5m telescope + Gunn filters + TI CCD.  However, the 
GISSEL code computes the color zero points using an A0 V star spectrum 
whereas the zero point of the Gunn photometric system is an F star.  
Therefore, an additional zero point is required to put the GISSEL 
$g-r$ colors on the standard Gunn system.  This zero point was 
calculated by taking the absolute fluxes of a GISSEL 14 Gyr E model 
and running them through a program which computes colors on the 
standard Gunn system.  The resulting additional zero point in $g-r$ is 
$-0.455$ mag, which is added to the GISSEL colors computed with the 
default filter definitions.  Because the $g$ filter used for the 
observations corresponds to the Thuan-Gunn $g$ band (which has a 
central wavelength $\sim 200$\AA~bluer than the ``standard'' Gunn $g$ 
transmission curve assumed by GISSEL), additional systematic color 
terms up to $\sim 0.1$ mag may be present in the $g-r$ colors.

For $q_0=0.5$ and $H_{\rm o}=50$~\kmps\/\Mpc$^{-1}$, $z=0.23$ 
corresponds to $t_0=10$~Gyr and to a look-back time of 3 Gyr.  The 
star-formation rate of a nominal elliptical galaxy (E) was modeled as 
a burst with a constant rate of star formation lasting for 1 Gyr.  The 
nominal spiral model (S) assumes a constant star-formation rate for 10 
Gyr.

In order to compare the colors of field galaxies with our cluster 
sample, GISSEL models were used to derive the following relative 
K-correction in order to convert an observed color $(g-r)_z$ at $z$ to 
$(g-r)$ at the fiducial redshift of $z=0.23$:
\begin{equation} (g-r) = (g-r)_z - 2.85 (z-0.23) .
\label{eqn:colorfix} \end{equation}

This correction is valid to within 0.05 mag in $g-r$ for all galaxy 
types between $0.17 \leq z \leq 0.37$, the range of our field 
galaxies.  To compare with work done on local galaxies, models were 
also used to derive the following relationship between $B-V$ measured 
in the rest frame and $g-r$ in the observed $z=0.23$ frame:
\begin{equation} (B-V) = 0.20 + 0.77 (g-r) \label{eqn:bvgr}
\end{equation} which is valid for both E and S SEDs, and 
\[ B =\left\{ \begin{array}{ll} 0.56 + 1.00 g + 0.63 (g-r), & \mbox{E SED}
\\ 0.52 + 1.04 g + 0.38 (g-r), & \mbox{S SED}. \end{array} \right. \]

\subsection{Checking the predictions of the GISSEL E model}

\begin{figure}
\centerline{\psfig{figure=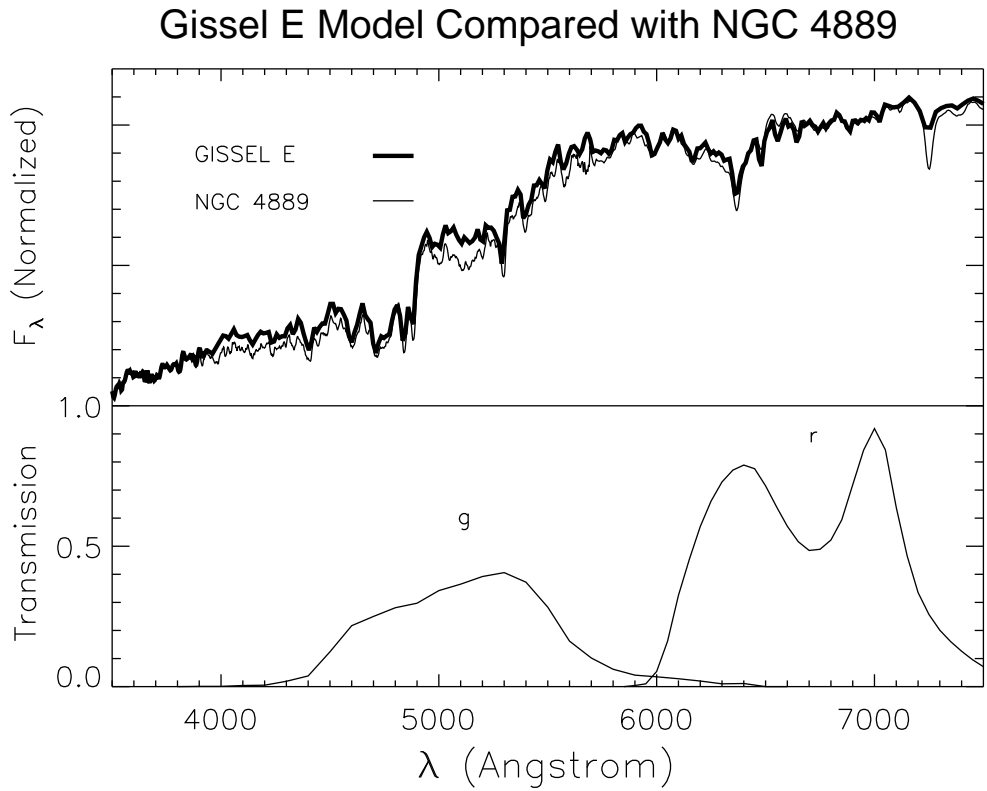,width=5in}}
 \caption{ Upper: Comparison of the GISSEL 14 Gyr E model
    (thick line) with a spectrum of NGC 4889 (thin line). Lower: the
    Gunn $g$ and $r$ filter transmission curves used by GISSEL.}
\label{ngc4889gisselE} \end{figure}

Unfortunately, the observed spectra in the present sample do not
extend to $r$ wavelengths, and hence these spectra cannot be used to
test the robustness of the predicted colors.  However, to check the
prediction of the GISSEL E model a synthetic 14 Gyr E spectrum was
compared with a high signal-to-noise spectrum of NGC~4889 (Oke, Gunn,
\& Hoessel 1995, in preparation), one of the brightest members of the
Coma cluster.  The comparison of the spectra redshifted to $z=0.23$ is
plotted in Figure~\ref{ngc4889gisselE}.  For reference, this figure
also shows the Gunn filter transmission curves used by GISSEL.  The
differences in the spectra are relatively small, and the computed
$g-r$ color of the model at $z=0.23$ is 0.08 mag {\it bluer} than the
NGC~4889 spectrum.  The observed NGC~4889 spectrum is extremely well
calibrated, but it was obtained through a narrow slit.  Because
galaxies do have internal gradients, a spectrum integrated over the
entire galaxy, as is the case for our A2390 data, may show somewhat
different spectral energy distribution.  Therefore, the GISSEL model
was also compared against an NGC~4889 spectrum taken through a wide
aperture \cite{Kennicutt:1992}.  Redshifted to $z=0.23$, the model is
0.09 mag {\it redder} in $g-r$ than the Kennicutt spectrum of
NGC~4889.  However, Kennicutt cautions that the data may have $\sim
10$\% uncertainties in the photometric calibration over the wavelength
range of interest.  The GISSEL models therefore appear to predict the
spectra and color of local elliptical galaxies fairly well, with
discrepancies in color $\simlt 0.10$ mag.

\section{COMPARIS0N OF DATA AND MODELS}

There are clear radial gradients in the colors, spectroscopic 
properties and morphologies of the galaxies in Abell 2390 
(Figure~\ref{fig:radpanel}).  Consequently the GISSEL models will be 
compared to the data in radial bins.  A successful model must account 
for the following key observations:

\begin{figure}
\centerline{\psfig{figure=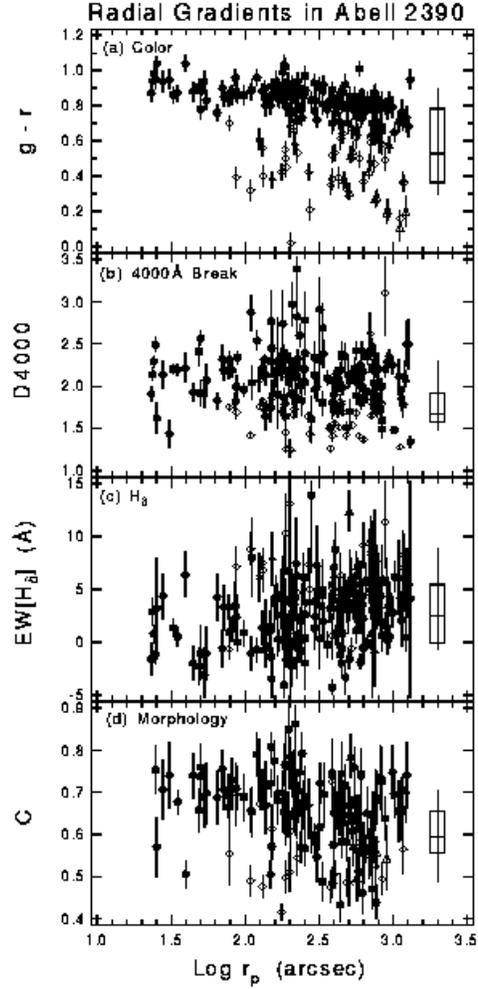,height=6in}}
 \caption{ The radial gradients in color, line measures,
    and morphology, plotted as a function of projected radius for our
    sample of cluster and field galaxies.  Filled circles are red
    cluster members, and open circles are blue cluster members.
    Similarly, filled triangles are red near-field galaxies, and open
    triangles are blue near-field objects.  Negative values of the
    line measures correspond to emission.  The symbol at the right of
    each panel is a ``Tukey box plot'' showing the field distribution.
    The box encloses the 25th and 75th percentiles for the field
    sample, and is subdivided by the median.  The vertical bar spans
    the range between the 10th and 90th percentiles of the field
    distribution.  Field galaxy colors have been K-corrected into the
    cluster rest frame.  See text for further details.}
\label{fig:radpanel} \end{figure}

(1) The almost complete absence of blue galaxies within the central 
$100\as$ of the cluster.

(2) The color gradient in the red galaxy sequence as a
function of radius.

(3) The relative increase in the blue fraction, and the increased 
blueness of the bluest galaxies, as a function of radius.

(4) The large radial extent of the dominant red galaxy sequence in the
cluster.

(5) The nature of the many cluster members seen at large radii whose 
colors and spectral indices are intermediate between those expected 
for E/S0s and normal spirals (eg. the HDS galaxies).

Explaining these observations with spectral synthesis models can be
rather complicated since age, metallicity and reddening may all be
important.  Color data alone are inadequate for testing models.  For
example, a model of an elliptical galaxy (constant star-formation rate
for $t \leq 1$ Gyr, solar metallicity) reddens by $\Delta(g-r) = 0.1$
between the age of $3.5$ and 10 Gyr \shortcite{Bruzual:1993}.  The
same amount of reddening is predicted for 10 Gyr old ellipticals if
the mean metallicity changes from [Fe/H] $ = 0$ to $+0.25$
\cite{Worthey:1994}. The 4000 \AA~break, D4000, can also be used as an
age indicator but it too is sensitive to metallicity in old stellar
populations. Although Dressler \& Schectman (1987) found only a weak
correlation between absolute magnitude (\ie metallicity) and D4000 in
E/S0 galaxies, more recent studies of field and cluster E/S0 galaxies
over a wider range of absolute magnitude find a significant
correlation (\shortciteNP{Kimble:1989}, \citeNP{Davidge:1994}).  In
their data, \shortciteN{Kimble:1989} find that D4000 correlates well
with a variety of metallicity indicators in the sense that galaxies
with smaller D4000 (fainter $M_B$) have weaker metal lines.  Although
D4000 is systematically lower for galaxies with [O~II] emission,
indicative of current star-formation, the bulk of the trend appears to
be more correlated with the strength of the metal absorption lines.
However, for the bright end of the luminosity function ($-22 \leq M_B
\leq -19$) the slope of the relationship between D4000 and $M_B$ is
relatively shallow (0.03 mag$^{-1}$).  Fainter than $M_B = -19$, the
relationship steepens, and hence metallicity has been inferred to be
the primary cause of D4000 variations in E/S0 galaxies fainter than
$M_{B}=-19$.

In the present work the star-formation history of cluster members is
constrained by using color and D4000 {\em in unison}, along with
measures of Balmer lines (which are mainly sensitive to age), in order
to model the observations.  These measures will be discussed
separately in the following subsections.  In the diagrams below the
temporal evolution of the E and S models (solid lines) are superposed
on the data.  Fiducial ages of $t=1$, 2, 3, 5, 8 and 10 Gyr are
indicated on the models with small dots.  To illustrate the radial
gradient in the galaxy population, the data has been split into 4 bins
defined by projected radius: (a) the inner cluster ($r_p \leq
100$\arcsec), (b) the cluster at intermediate radii ($100\as < r_p
\leq 400\as$), (c) the cluster at outer radii ($r_p > 400\as$), and
(d) the field.  The colors of the field sample have been k-corrected
to $z=0.23$ using equation~\ref{eqn:colorfix}.  Galaxies with [O II]
emission detections $>2\sigma$ are plotted as open circles.

\subsection{Color and 4000 \AA\/ break}

\begin{figure} 
\centerline{\psfig{figure=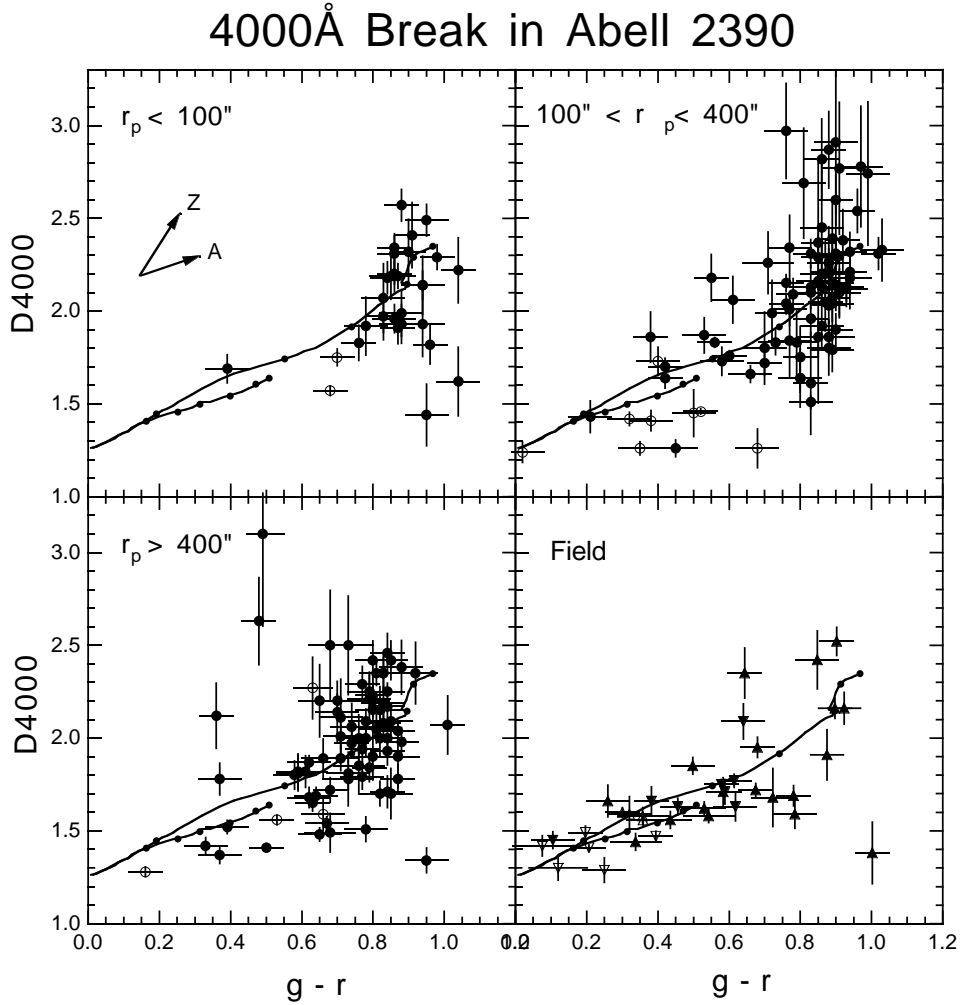,width=5in}}
\caption{ D4000 vs color for galaxies in the cluster and
    field for different bins of projected radius.  The Z vectors show
    the effects of changing [Fe/H] in an old elliptical from
    $0$~to~$+0.25$.  The A vector shows the change produced by a
    reddening of $E(g-r) = 0.18$ mag ($A_V=0.5$ mag) in the rest frame
    of the cluster for the elliptical model (see text for details).
    Open symbols denote galaxies with $2\sigma$ detections of [O II]
    emission. In the field galaxy panel redshifts have been limited to
    $0.17<z<0.39$ in order to correspond to the regime where
    K-corrections can be estimated accurately.  Upward-facing
    triangles are field objects, and downward facing triangles are
    near-field objects.  The temporal evolution of the E and S models
    are shown as solid lines, and ages $t=1,$ 2, 3, 5, 8 and 10 Gyr
    are identified with small filled circles.}
\label{fig-d4000gr} \end{figure}

\begin{figure}
\centerline{\psfig{figure=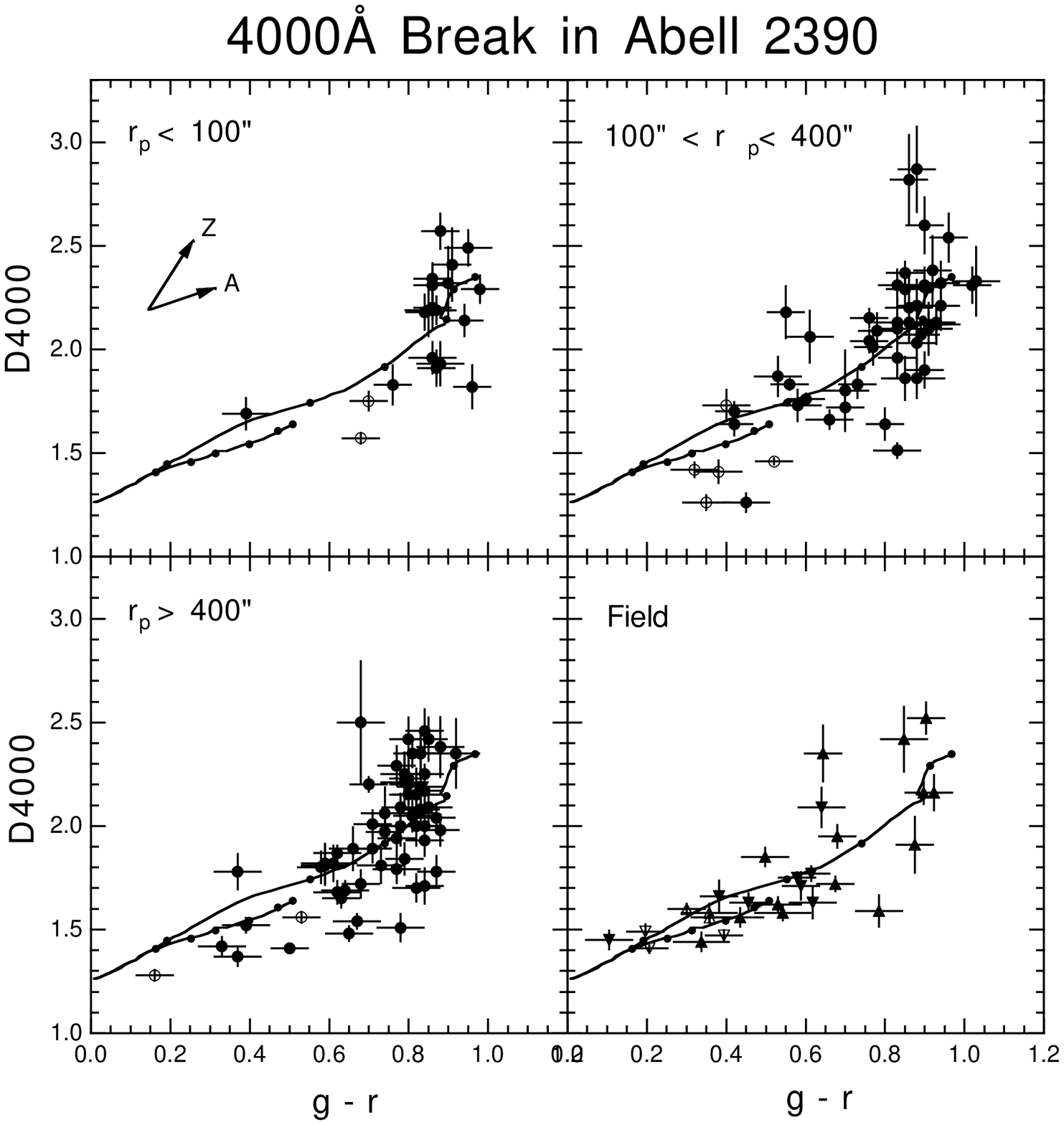,width=5in}}

 \caption{ As for the previous figure, except that the
    sample has been restricted to objects with high signal-to-noise
    level (${\rm SNR}>6$) spectra.}
\label{fig-d4000grSNR} \end{figure}

Figures~\ref{fig-d4000gr} and \ref{fig-d4000grSNR} show D4000 versus
color, both for our complete sample and for a subsample of galaxies
with high signal-to-noise spectra (SNR$>6$).  The vector labeled A
illustrates the change produced by a reddening of $E(g-r) = 0.18$ mag
($A_V=0.5$ mag) in the rest frame of the cluster for the E model.  The
vector labeled Z shows the movement of 10 Gyr old elliptical if
metallicity varies from [Fe/H] $= 0$ to $+0.25$ \cite{Worthey:1994}.

\subsubsection{Inner Cluster Members {\rm (}$r_{p}<100\as${\rm )} }

The inner cluster galaxies are in good agreement with the GISSEL 
models, and have a small color dispersion but moderate dispersion in 
D4000.  This may be caused by either age or metallicity variations.  
As can be seen from the models, changes in age and metallicity fall 
along roughly parallel lines, and hence they are inseparable in this 
plane\footnote{However, it is interesting to note that a co-added 
spectrum of the four innermost cluster galaxies shown in 
Figure~\ref{fig-d4000gr} with D4000 $\approx 1.9$ shows a higher 
H$\delta=3$~\AA\/ and weaker Ca II K than the co-added spectrum of the 
inner 12 galaxies with D4000 $>2$.  These could be signs of age rather 
than metallicity differences, but higher S/N data are required in 
order to investigate this further.}.  In the absence of synchronizing 
effects, the narrow color dispersion envelope at small radii implies 
that the galaxies in the inner parts of the cluster have a relatively 
small age spread
\shortcite{Aragon-Salamanca:1993}.  We can definitively say that
these galaxies have had little star formation in the 3 Gyr prior to 
the epoch of observation, and only four galaxies can be significantly 
younger than 5 Gyr.  Assuming that their epoch of star formation 
lasted $\sim1$ Gyr and their mean metallicities are approximately 
solar, the remaining galaxies are consistent with being coeval with 
ages $=8\pm 2$ Gyr at $z=0.23$.  Thus their redshift of formation is 
large, $z_f\ge2$, and they must have formed early during the initial 
collapse of the cluster.  Although the dispersion in D4000 that is 
seen in the central galaxies is larger than expected based on their 
absolute magnitude range, we cannot conclude that this represents a 
dispersion in age or metallicity because we may have additional 
systematic errors in D4000 due to uncertainties in flux calibration of 
the slitlets.

\subsubsection{Outer Cluster Members {\rm (}$r_{p}>100\as${\rm )}}

Outside $r_{p}=100\as$ the models remain in good agreement with the 
D4000 and color data.  The galaxies occupy the regions predicted for 
star-forming galaxies (S model), and along the passively evolving E 
model.  {\em The galaxies apparently track an age sequence}, and the 
agreement with the models is excellent because D4000 is much more 
sensitive to age than to metallicity for young stellar populations.  
Galaxies bluer than $g-r=0.4$ lie near or on the S track and have 
active star-formation (low D4000, [O II] emission).  Objects with 
$0.5<g-r<0.7$ correspond to systems in which star-formation has ceased 
(we will show in the next section that these systems generally have 
strong Balmer absorption), and consequently D4000 is higher.  The 
reddest galaxies correspond to progressively older, passively evolving 
populations and occupy the same position in the D4000-color plane as 
the inner objects with $r_{p}<100\as$.  Outside the cluster, the field 
population is in excellent agreement with the predictions of the model 
(90\% of the galaxies lie on or near the S track or the young part of 
the E track).

\subsection{H$\delta$ Evolution}

\begin{figure}
\centerline{\psfig{figure=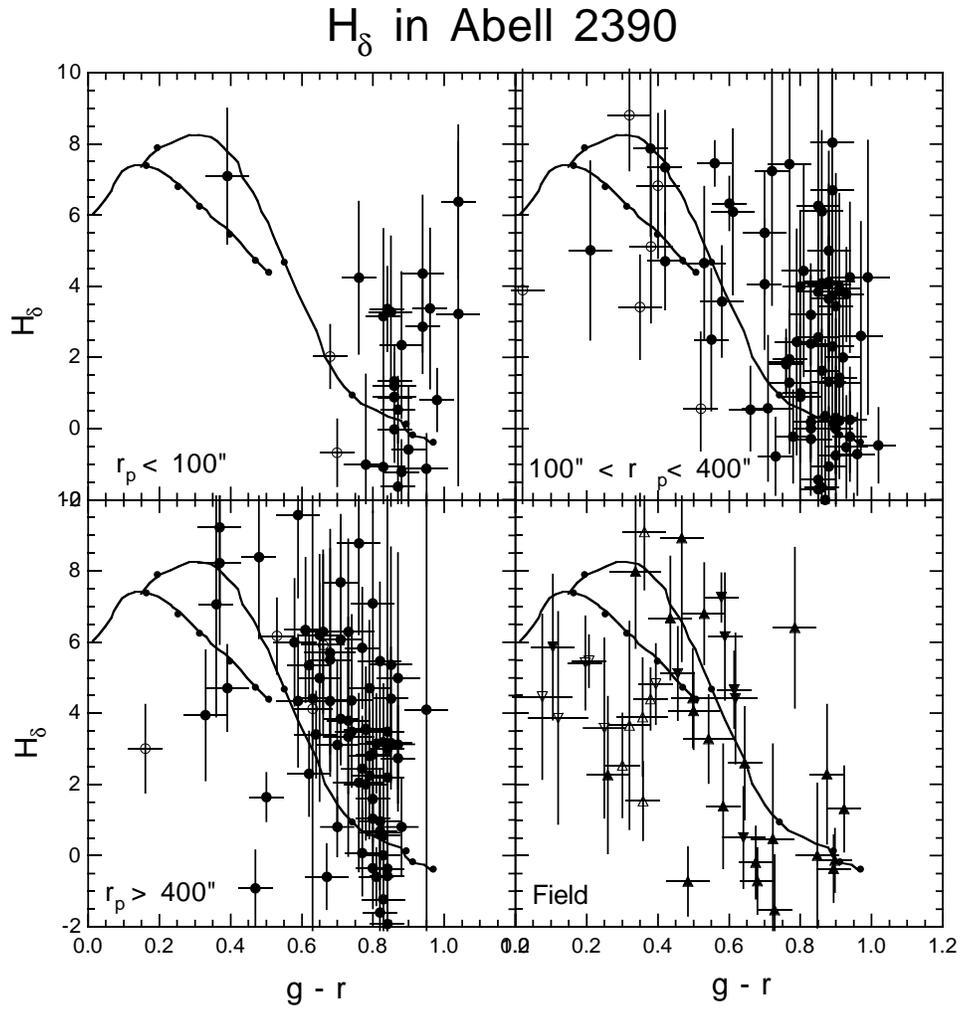,width=5in}}

\caption{ Equivalent width of H$\delta$ vs color for
    (a--c) cluster members in bins of projected radius $r_p$ (as for
    previous figure), and (d) field galaxies.  Solid lines are the
    nominal E and S models.  The plot symbols have the same meaning as
    for the previous figure.}
\label{fig-hdeltarp} 
\end{figure}

\begin{figure}
\centerline{\psfig{figure=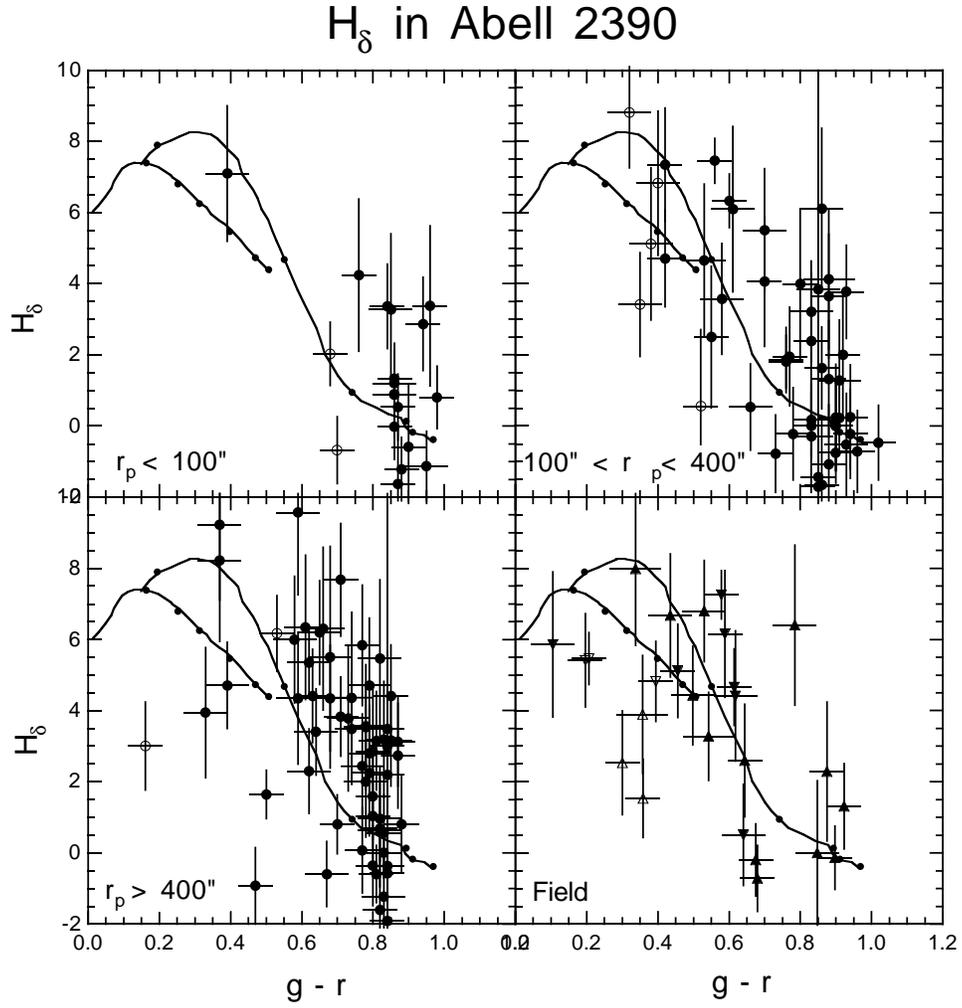,width=5in}}
\caption{ As for the previous figure, except that the
    sample has been restricted to objects with high signal-to-noise
    level (${\rm SNR}>6$) spectra.}
\label{fig-hdeltarpSNR} \end{figure}

In Figures~\ref{fig-hdeltarp} and~\ref{fig-hdeltarpSNR} H$\delta$ vs
color is shown using same bins of projected radius adopted for the
D4000 plots.  As with the previous figures, the nominal E and S models
have been superposed on the data.  There is excellent agreement
between the models and observations for galaxies in the field.  The
agreement is also good for objects with radii \( r_{p}<100\as \).
This is not true for objects at intermediate radii (\(
r_{p}>100\as\)): either H$\delta$ is too strong, galaxies are too red,
or some combination of both of these effects is occurring.  These
objects (the most striking of which have already been discussed in
Section 4.3) apparently correspond to the HDS (and E+A) systems found
by \citeN{Dressler:1983} and \citeN{Couch:1987} in their cluster
surveys.

\begin{figure}
\centerline{\psfig{figure=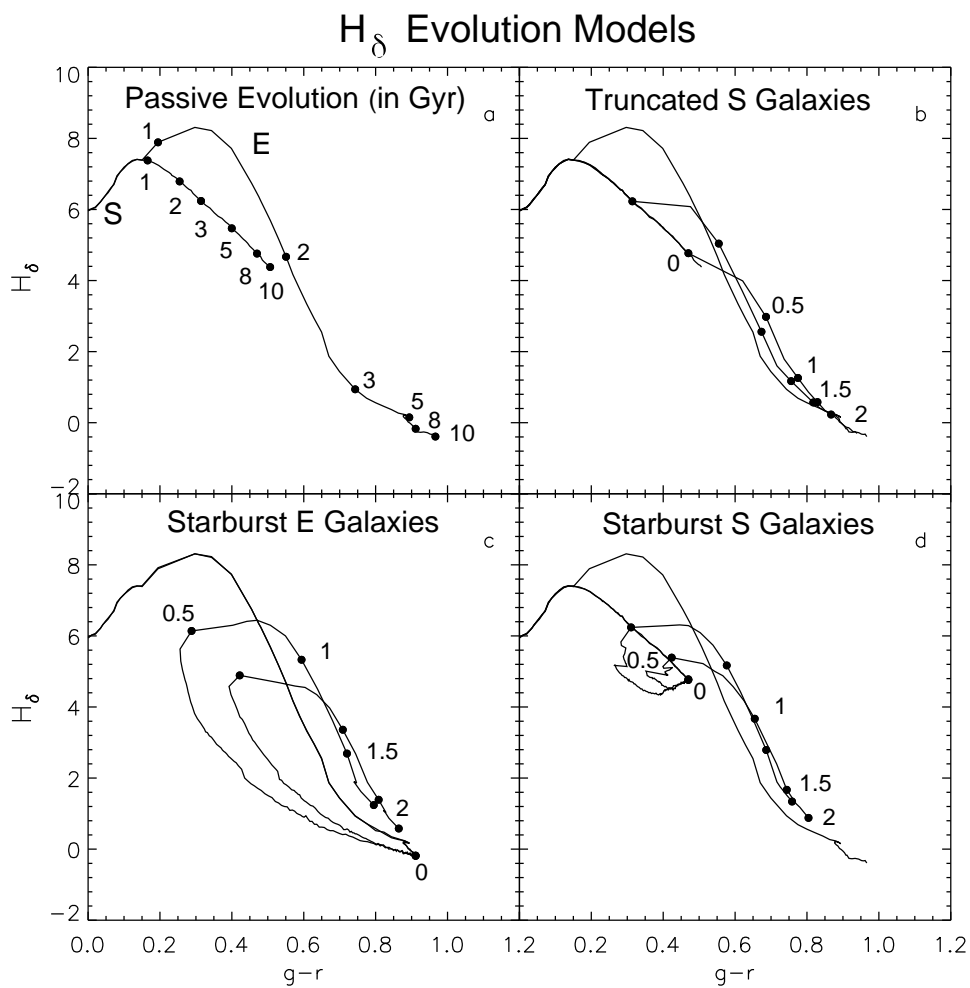,width=5in}}

 \caption{ Models of H$\delta$ evolution: (a) Nominal E and
    S model. The numbers correspond to age in Gyr. (b) Truncated S
    models. The numbers correspond to time (in Gyr) after the onset of
    truncation. (c) star bursts in the E model, and d) star bursts in
    the S model. Numbers correspond to time (in Gyr) after the burst.
    See text for more details.} \label{hdelmod1} \end{figure}

\begin{figure} 
\centerline{\psfig{figure=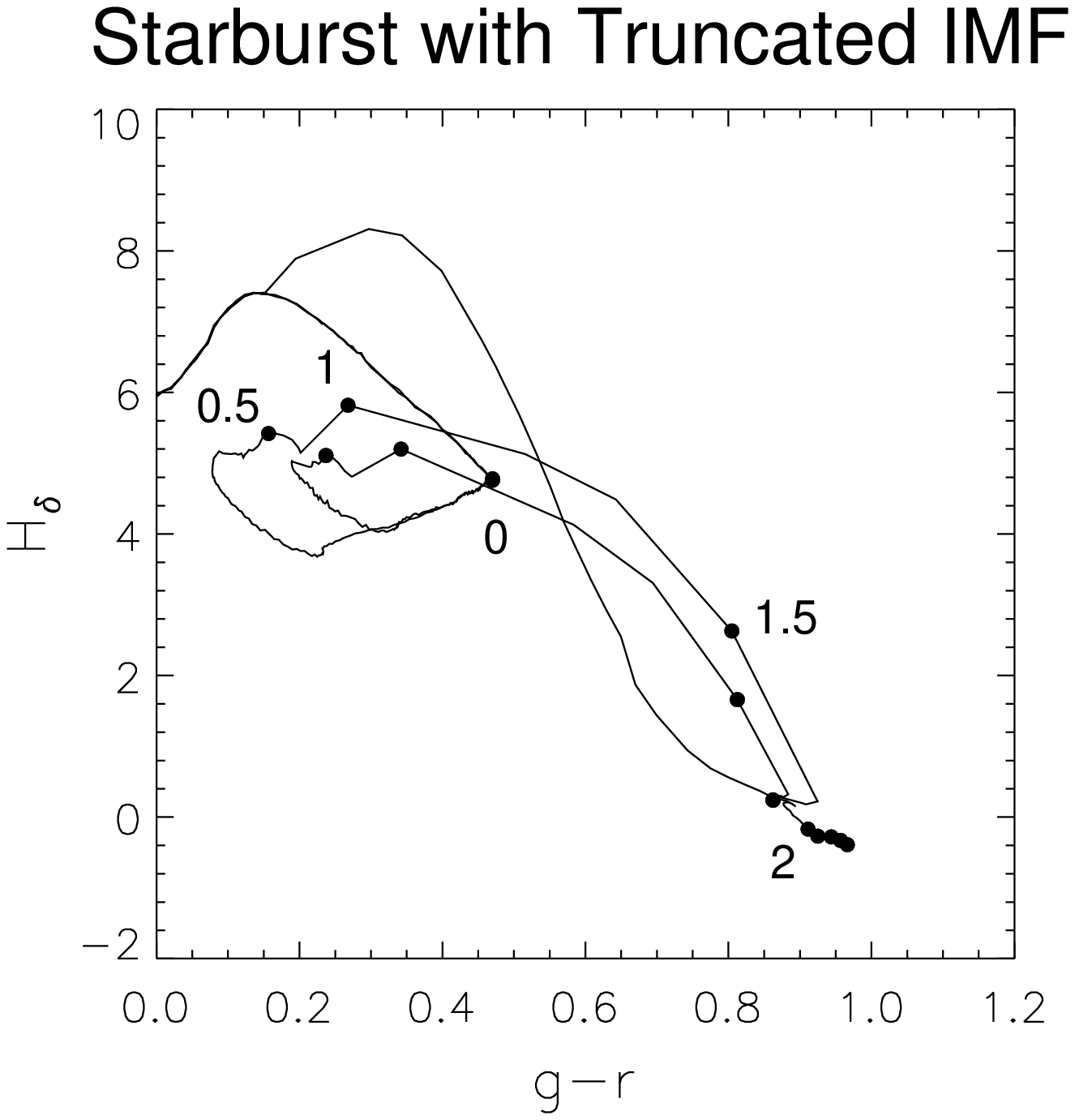,width=5in}}

\caption{ Model of H$\delta$ evolution: a star burst in
    the nominal S model with a truncated IMF that forms only massive
    stars (2.5 to 125 $\msun$). Numbers on the tracks correspond to
    time (in Gyr) after the burst. }\label{hdelmod2} \end{figure}

In order to investigate the nature of these HDS systems, GISSEL was
used to compute the evolutionary tracks of both starburst and
truncated star formation models on the H$\delta$-color diagram
(Figure~\ref{hdelmod1}).  In each panel of this Figure the nominal E
and S models have been displayed as a benchmark.  In panel (a) the
positions corresponding to passively evolving galaxies with ages
$t=1$, 2, 3, 5, 8 and 10 Gyr are indicated.  Panel (b) shows shows two
truncated star-formation models in which the nominal S is truncated at
$T=3$ and 8 Gyr, and ages $t = T$ + [0, 0.5, 1, 1.5, 2] Gyr are
marked.  Panel (c) shows two models of star bursts in an E galaxy
which form 10 and 20\% of the total mass of stars formed in the
galaxy.  The bursts are assumed to begin at $T=8$ Gyr and last 1 Gyr.
The ages marked are the same as in (b).  Panel (d) shows the identical
two bursts in an S galaxy.  {\em In each starburst model, we assume
that star formation ceases after the burst.} Figure~\ref{hdelmod2}
shows the same S + starburst model under the assumption that IMF in
the burst is biased to form only massive stars (2.5 to 125 $\msun$).
The observed distribution of HDS objects on the $H\delta$-color
diagram is in qualitatively good agreement with the evolutionary
tracks shown in Figures~\ref{hdelmod1} and ~\ref{hdelmod2}. However,
the observed $H\delta$ values tend to be somewhat larger than
predicted by the both starburst and simple truncated star formation
spectral synthesis models.  Models with IMFs biased toward massive
stars reproduce a steep H$\delta$ vs $g-r$ as seen in the data, but
the steepness of the H$\delta$ points may also be produced by factors
such as temporal variations in internal reddening in truncated or
burst models, or by changes in metallicity. These factors are not
accounted for in the GISSEL models.

It is apparent from Figures~\ref{hdelmod1} and~\ref{hdelmod2} that the
behavior of $H\delta$ can be qualitatively understood as the
manifestation of a {\em halt in star formation}.  The key question to
be addressed is whether or not this halt is due to (a) the resumption
of ordinary behavior in ellipticals that have undergone bursts (the
original E+A scenario proposed by Dressler \& Gunn 1983), (b)
exhaustion in the gas supply of disk systems that have undergone
starbursts (favored by \citeNP{Couch:1987}), or (c) truncation in the
star formation rates of ordinary (i.e.  non-bursting) disk systems.
However, once galaxies have reached the HDS stage the shapes of the
evolutionary tracks for all these scenarios are rather similar.  In
the starburst model H$\delta$ (as well as color and D4000) declines on
a timescale $\sim 2$ Gyr after the burst has ceased, irregardless of
whether the initial galaxy is a S or E.  After the burst, the E +
burst model is only slightly redder than the the S + burst model.
Given the relatively large uncertainty in our $H\delta$ measurements
the later evolutionary stages of the burst and truncated
star-formation tracks are effectively indistinguishable.  Therefore
the key observable quantity in our data that can be used to
discriminate between starburst and simple truncated star formation
scenarios is the fraction of galaxies at early points on the
evolutionary tracks versus the number of potential post-starburst
systems (ie. the number of systems currently undergoing starbursts
versus the number of HDS galaxies).  

\subsection{The starbursting fraction}

\begin{figure} 
\centerline{\psfig{figure=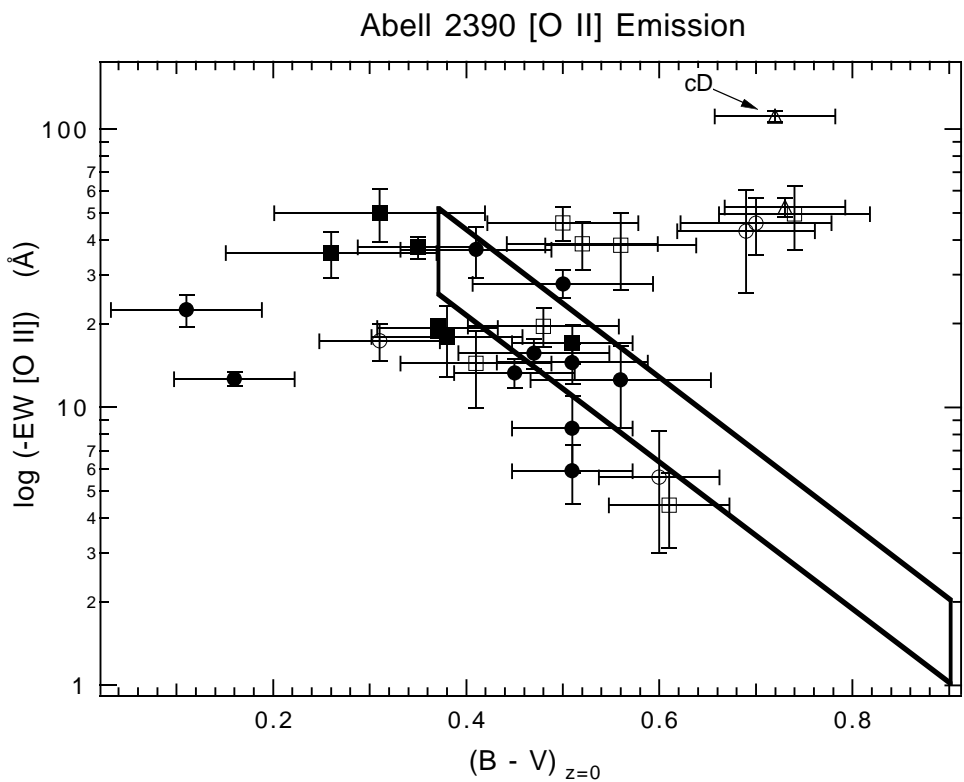,width=5in}}
\caption{ The equivalent widths of [O~II] 3727 in galaxies
    with $>2\sigma$ detections: triangles are inner cluster members
    ($r_p \leq 100$), squares are cluster members at intermediate
    radii ($100 < r_p \leq 400$), circles are cluster members at large
    radii ($r_p > 400$), filled squares are near-field galaxies and
    filled circles are field galaxies.  The trapezoid represents the
    area occupied by a sample of nearby cluster and field galaxies
    Dressler (1982).}
\label{fig:o2detns}
\end{figure}

The equivalent widths and rest-frame colors of cluster and near-field
galaxies with [O II] emission detected at the 2$\sigma$ level are
shown in Figure~\ref{fig:o2detns}.  Equation~\ref{eqn:bvgr} has been
used to convert observed $g-r$ to rest frame $B-V$.  The trapezoid
superposed on the points corresponds to the region occupied by nearby
cluster and field galaxies with ordinary emission line characteristics
\cite{Dressler:1982}.  In local galaxy samples, Sbcs typically have
$B-V\approx 0.58$ mag and [O~II] $\approx 10$\AA, Scds have
$B-V\approx 0.48$ mag and [O~II] $\approx 17$\AA, and Ims have
$B-V\approx 0.48$ mag and [O~II] $\approx 48$\AA~(Coleman, Wu \&
Weedman 1980, Kennicutt 1992).  In the Abell 2390 sample of fourteen
cluster members with detected [O~II] emission at the $2\sigma$ level,
six galaxies appear to be normal spirals, three are either normal
spirals or galaxies with star formation rates slightly (a factor of
$\sim 2$) higher than normal, and five objects (only 2\% of the total
cluster population) are starburst/AGN candidates. Amongst the six
near-field [O~II] emitters, five objects are consistent with being
late-type systems, and one galaxy is probably a starburst. The
properties of the individual starbursting systems in our sample are
described in \S7.4.

\section{Discussion}

\subsection{Truncated star formation}

On the basis of Figures~\ref{fig-hdeltarp} -- \ref{fig:o2detns} one
can immediately rule out the possibility that galaxy evolution in
clusters is a {\em continuous} process proceeding via starbursts.  Our
observations rule out the possibility that large numbers of starbursts
are currently occurring.  Furthermore, the relative number of galaxies
at different positions along starburst model evolutionary tracks in
Figure~\ref{hdelmod1} rules out the possibility that {\em steady}
evolution in the galaxy population occurred via starbursts in the
recent past.  Figure~\ref{hdelmod1} indicates that both bursting
ellipticals and bursting spirals spend 40--45\% of the first 1.5 Gyr
following a starburst blueward of the passive evolution elliptical
curve before looping across and joining the elliptical curve $\simgt
2$ Gyr after the burst.  However, in the data shown in
Figures~\ref{fig-hdeltarp} and~\ref{fig-hdeltarpSNR}~only 15--20\% of the
galaxies lie blueward of the passive evolution track in the region of
the diagram corresponding to times between 0 and 1.5 Gyr following a
burst.  Therefore the majority of the HDS population must have evolved
via truncated star formation {\em that did not follow from an initial
starburst}.  Furthermore, because of the selection criteria in the
present sample, the fraction of non-burst HDS galaxies determined on
the basis of relative position along the tracks in
Figure~\ref{hdelmod1} is likely to be significantly underestimated.
Bursting galaxies (and post-starburst galaxies blueward of the passive
evolution tracks) have been brightened, while HDS objects redward of
the passive evolution tracks have faded.  The magnitude selection
function for the present sample is falling rapidly near $r=21$ mag,
and intrinsically less luminous objects are under-represented.

The argument given above suggests that if wholesale changes in the 
cluster galaxy population have been precipitated by starbursts, then 
such changes must have been episodic, with the last period of 
starbursts ending about 1 Gyr before the epoch of observation.  
Episodic bursts are difficult to constrain from spectral synthesis 
models, since one can always argue that at the epoch of observation 
the cluster is no longer evolving in the same manner as in the past.  
However, if the galaxies in Abell 2390 form a radial sequence as we 
postulate, and if episodic starbursts are triggered by specific 
environmental conditions, then it seems reasonable to argue that at 
{\em some} radius galaxy evolution via starbursts should be occurring.  
The large radial coverage in the present dataset effectively rules 
this out.  Furthermore, if episodic starbursts (rather than steady 
evolution) dominate the evolutionary process in clusters, then one 
expects to see a large scatter in the cluster blue fraction as a 
function of redshift.  Both the results of
\citeN{Butcher:1984} and preliminary results from the CNOC survey 
(\citeNP{Yee:1995}) suggest a steady monotonic rise in the cluster 
blue fraction as a function of redshift.

It is therefore suggested that galaxy evolution in Abell 2390 is being
driven by the {\em steady truncation in the star formation rate of
infalling galaxies}.  In this scenario Abell 2390 is composed of a
central red old population, and an outer envelope of progressively
younger galaxies which have been accreted from the field over $\sim8$
Gyr.  It is important to be clear about what is meant by ``younger''
galaxies in this context, since age can be measured either in terms of
time since the initial onset of star formation or in terms of the mean
age of the stellar population.  At a given radius the slope of the red
galaxy locus on the color-radius diagram implies that (on average) the
{\em oldest} galaxies at that radius {\em ceased forming stars} more
recently than did the oldest galaxies nearer to the center of the
cluster.  However star-formation models do not give unique
predictions.  One can produce the bluer colors either by starting star
formation in the galaxies at later times, having an increase in the
star-formation rate at later times, or truncating a constant
star-formation rate at later times.  We interpret the increased
dispersion in the colors of blue galaxies on the color-radius diagram
to mean that at larger cluster radii galaxies have had longer periods
of star formation before that star formation was halted by infall into
the cluster.  Hence our claim is also that the {\em the mean age of
the stellar population in galaxies becomes younger with increased
radius in the cluster}.  This dynamical picture of a growing cluster
is consistent with hierarchical scenarios for the evolution of large
scale structure \cite{Gunn:1972}.

This scenario is also consistent with the remarkable East/West
dichotomy in the distribution of line emitters that is seen in
Figure~\ref{fig-spatial2}, and with the bulk motion along the West
side of the cluster indicated by the Dressler-Shectman test shown in
Figure~\ref{fig:spatial} (and roughly centered on the NW Group).
These suggest that Abell 2390 is accreting objects from the field
non-isotropically, perhaps along a ``sheet'' of large scale structure
joined to the West side of the cluster. Some infalling objects may
undergo mild starbursts induced by interactions with the ICM, but
Figure~\ref{fig:o2detns} suggests that most infalling line emitters
are simply late-type systems common in the field. Assuming infalling
objects have peculiar motions of order $\sim1000$~km~s$^{-1}$,
accreted galaxies in the periphery of the cluster move one~Megaparsec
in a Gigayear (the approximate timescale for fading the disk of a
field galaxy and turning it into a red HDS system). Since infalling
objects fade before completing a single cluster crossing, the
spatially skewed color distribution between the cluster and infalling
field population on the West side of the cluster is preserved.

\begin{figure}  
\centerline{\psfig{figure=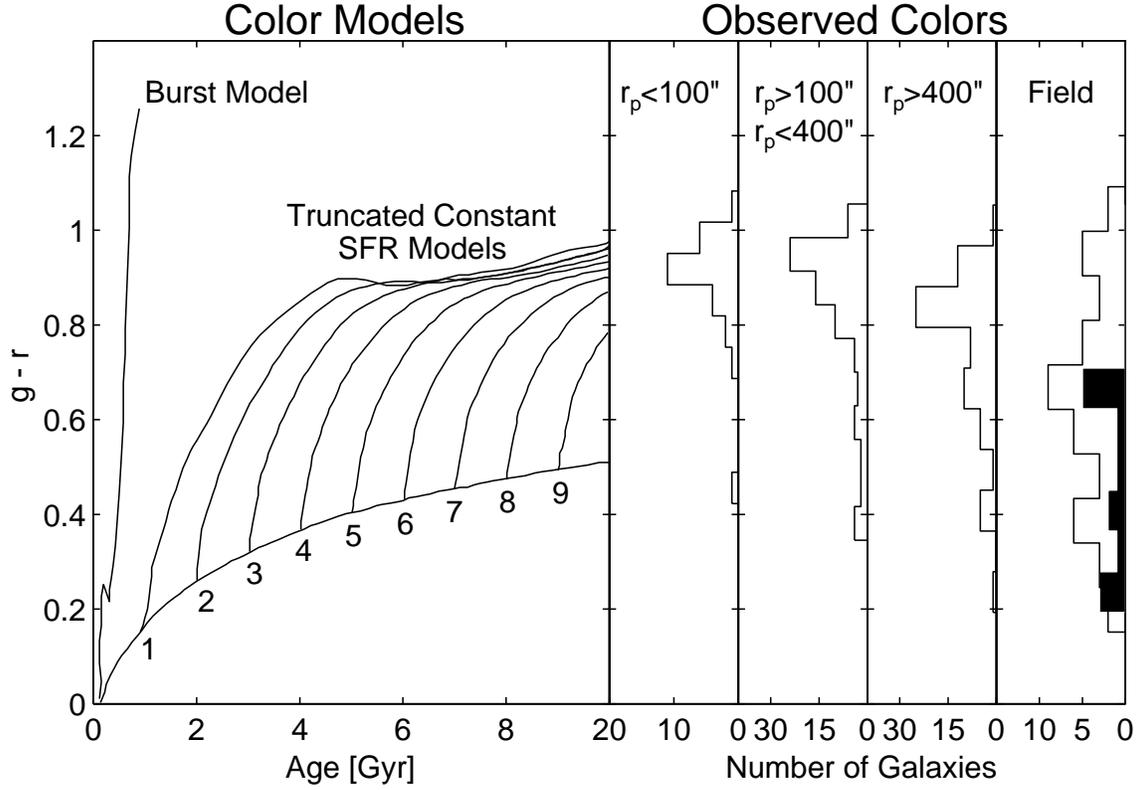,width=6in}}
\caption{Illustration of the agreement between the 
    color models and the observed colors at different radii.  [Left]
    the color-age relation for truncated, constant star-formation rate
    models.  The models are truncated at various ages from 1 to 10 Gyr
    in 1 Gyr intervals.  The short line which peaks at a very red
    color is a burst model assuming an IMF which is biased to form
    only massive stars (2.5 to 125 $\msun$).  Note that the stars have
    all evolved to near-maximum redness by $t \sim 1.5$ Gyr after
    truncation.  [Right] histograms of the colors of galaxies which
    are (from left to right) inner cluster members, cluster members at
    intermediate radii, outer cluster members, and field galaxies.
    The galaxies shown in grey are the near-field galaxies.}
    \label{fig-trunc}
\end{figure}

It is illustrative to compare directly the observed colors of cluster
galaxies with predicted colors as a function of age and duration of
star formation prior to truncation (Figure~\ref{fig-trunc}).  The
$g-r$ colors calculated from the GISSEL package are displayed on the
left half of this figure, under the assumption that truncated
star-formation has occurred in Abell 2390.  The lines, from left to
right are for star-formation durations of 1~Gyr to 10~Gyr, at
intervals of 1~Gyr.  On the right half of Figure~\ref{fig-trunc}
histograms indicate the observed color composition in the cluster as a
function of radius.  The radial bins are identical to those in
Figure~\ref{fig-d4000gr} (a--d from left to right).  With a few
exceptions, the colors fall into the range predicted by the model.
The two galaxies (1\% of the total) with bluer colors are entirely
consistent with being chance projections of field galaxies.  While
there are clearly {\em some} strong [O~II] emission line objects in
the present cluster sample, these are consistent with being
comparatively ``minor'' starbursts (resulting from the continuous
formation of $\simlt 20\%$ of the galaxy), and probably do not play a
major role in driving cluster galaxy evolution.

\subsection{Morphological Evolution}

The stellar populations (as measured from colors and spectroscopic
features) of the cluster members are consistent with a truncated star
formation scenario, but a self-consistent model for galaxy evolution
in Abell 2390 must not only account for colors and line features, but
also for morphological composition.  Truncation in the star formation
rate must account for a gradual radial evolution in the morphologies
of the galaxy population, starting from the late-type-dominated field
population, through ``anemic'' spirals, and leading ultimately to S0
galaxies.  Truncated star-formation models for the origin of S0s have
been proposed many times, and diverse physical mechanisms have been
proposed for halting star formation, such as ram-pressure stripping
\cite{Gunn:1972} and gas evaporation \cite{Cowie:1977}. However, the
``Nature vs. Nurture'' formation history of S0s remains controversial
(see \citeNP{Haynes:1988} for a review).  A major objection to an
evolutionary linkage between spirals and S0's appears to have been
eliminated by the work by \citeN{Solanes:1989} showning that the
increase in the disk-to-bulge ratio of spirals in denser regions (seen
in the Dressler [1980] data) is due to a diminution of disk
luminosity, rather than an increase in bulge luminosity as originally
thought.

To examine the effects of truncated star formation on morphology, 
GISSEL models were used to compute separately the expected fading in 
the disk and bulge components of late-type systems at $z\sim 0.25$.  
As before, disks were modeled using a constant star-formation rate.  
Bulges were modeled in the same manner as for elliptical galaxies (a 
constant star-formation rate for 1 Gyr).  Assuming a Scalo (1986) IMF 
and an age for the composite system of 7 Gyr, one Gyr after truncating 
the star formation a typical disk has faded by $\sim 1$ mag, while the 
bulge component has faded by at most a few tenths of a magnitude.  
Thus the bulge-to-disk ratio increases by a factor of $\gtrsim 2$ 
after one gigayear, increasing $C$ by 0.1 --- 0.2 for late-type 
systems (see Fig.1(b) in Abraham et al [1994]).

To illustrate the impact of this effect upon the radial $C$
distribution in the cluster, a very simple model was constructed for
the morphological composition of a cluster made up of an old component
and an infalling field population (whose star formation is truncated
by ingress into the cluster).  The old component was assumed to be
composed of ellipticals and S0s.  The infalling field population was
assumed to be comprised of 20\% ellipticals, 20\% S0 galaxies, 20\% Sa
galaxies, 20\% Sb galaxies, and 20\% Sc galaxies, which is roughly
appropriate for an $r$-selected sample \cite{Yee:1987}.  An even more
basic assumption was that the Hubble types could be modeled by the
superposition of an exponential disk and a de Vaucouleurs law bulge,
with bulge-to-disk ratios for these Hubble types typical of those seen
at low redshift.  The scale lengths and characteristic surface
brightnesses of the disk and bulge components were determined using
the relations given in \shortciteN{Strom:1988} and
\shortciteN{Boroson:1981}, assuming that the scale length for the
bulge component of spirals follows the canonical relation for
ellipticals, as suggested by \citeN{Kormendy:1985}.  Differential
K-corrections for the bulges and disks were applied using the spectral
energy distributions in \citeN{Coleman:1980} prior to fading the
components.

\begin{figure} 
\centerline{\psfig{figure=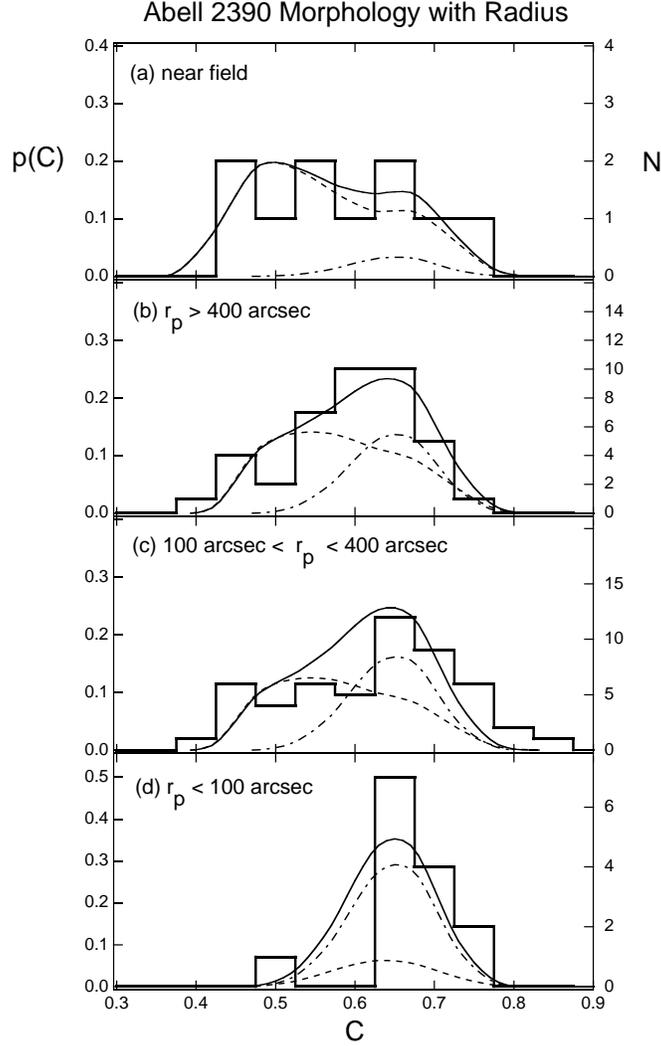,width=3.4in}}
\caption{ Histograms showing the distribution of
    morphological parameter $C$, binned as a function of radius.
    Superposed are curves corresponding to the simple model described
    in the text. The solid line is the superposition of the $C$
    distributions for a faded-disk field model (dashed line) and an
    old elliptical model (dot-dashed line). In the near field bin (a),
    our models assume that 90 \% of the galaxy population is made up
    of the faded field population (with a disk fading of 0.2 mag), and
    that the remaining 10\% of the galaxies are old ellipticals. In
    panel (b) we assume disks in the field population have faded by a
    mean value of 0.5 mag, and that the faded field population
    comprises 75 percent of the cluster population. In panel (c) we
    assume a mean fading of 0.7 mag, and that the faded field
    population comprises 50 percent of the cluster population. In the
    innermost bin (d), we assume the disks in the field population
    have faded by 1.5 mag, and that old ellipticals comprise 80
    percent of the galaxy population.}
    \label{fig:obsmorphbins} 
\end{figure}

The fraction of old early type systems (the relics from the initial 
cluster collapse) varied from 90 percent of the galaxy population in 
the center of the cluster to 10 percent at 3 Mpc from the cluster 
core.  The disk population was linearly faded as a function of radius, 
with no fading in the cluster periphery and 1.5 mag fading in the core 
of the cluster.  Figure~\ref{fig:obsmorphbins} shows the morphological 
data for Abell 2390, binned in radius, with the results from our 
simple model superposed.  The simplistic truncated star-formation 
model predicts a radial distribution of $C$ that is in qualitatively 
good agreement with our morphological data.  It therefore appears that 
the morphological composition of the cluster is consistent with the 
suggestion from the colors and line indices that the gradients seen in 
the cluster are the result of an age sequence in an infalling field 
population whose star formation has been halted by entry into the 
cluster.

\subsection{The Butcher-Oemler Effect}

\begin{figure} 
\centerline{\psfig{figure=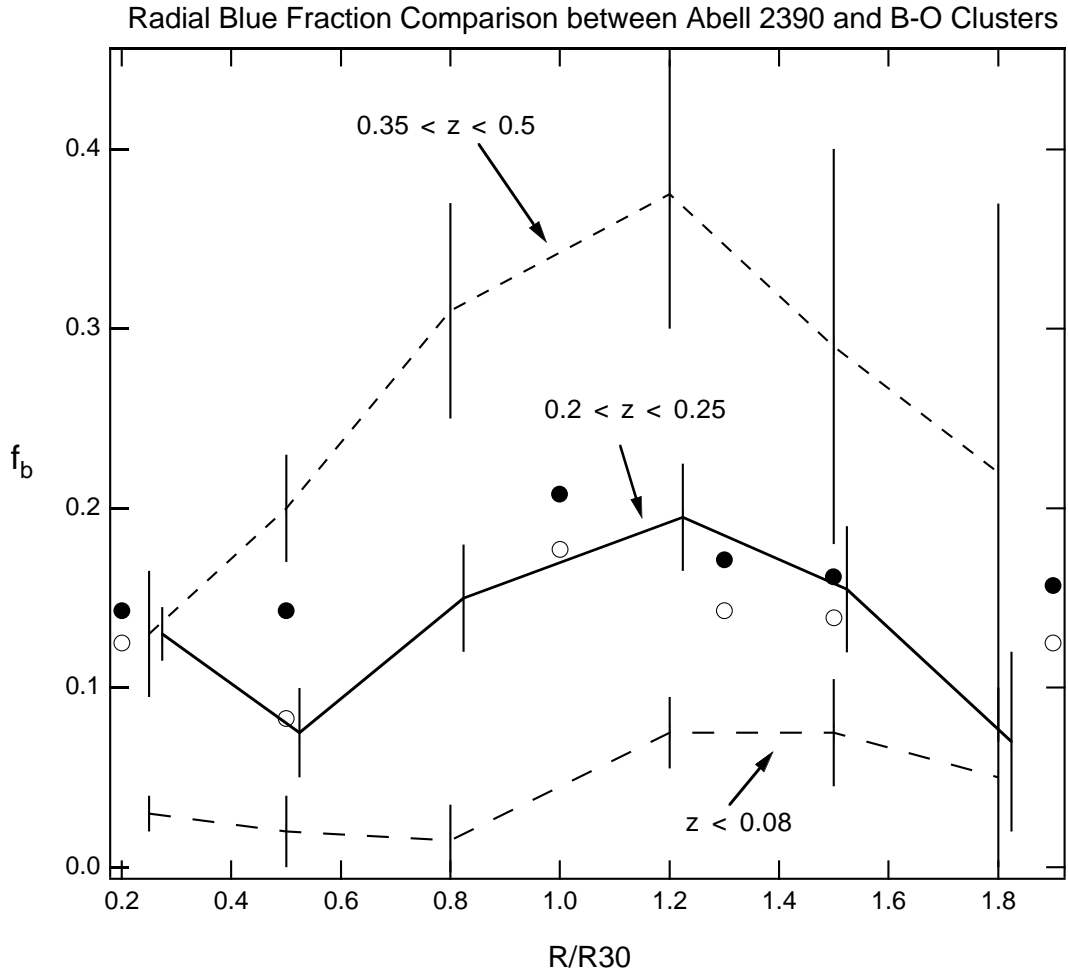,width=5in}}
\caption{ A comparison between the radial blue fraction in
    Abell 2390 (points) and the low redshift (lower dashed line),
    intermediate redshift (middle dashed line), and high redshift
    (upper dashed line) samples of Butcher \& Oemler (1984).  The
    Abell 2390 data is shown for two limiting magnitudes, $r<20.5$
    (open circles) and $r<20$ (filled circles), in order to
    demonstrate the low sensitivity of the blue fraction to the choice
    of limiting magnitude.}
    \label{fig:bocomp}
\end{figure}

It is interesting to consider whether the truncated star formation 
scenario proposed here may be relevant to galaxy evolution in other 
rich clusters, as characterized by the Butcher-Oemler effect.  In 
order to determine this, the blue fraction in Abell 2390, $f_b$, was 
measured using the same prescription adopted by \citeN{Butcher:1984} 
in their classic paper.  The number density profile of the cluster was 
calculated (assuming circular symmetry) and this profile was 
integrated in order to determine $R_{30}$, the radius within which 30 
percent of the cluster light was enclosed.  $R_{60}$ and $R_{20}$ were 
also calculated in order to determine the cluster concentration 
parameter $C_{BO}$, given by $C_{BO}=\log\, R_{60}/R_{20}$, so that 
Abell 2390 could be compared to Butcher-Oemler clusters of similar 
concentration.  Colors were normalized to $r=19$ mag and transformed to 
$B-V$ using Equation \ref{eqn:bvgr} in order to to reproduce what was 
done for the B-O clusters.  The blue fraction was determined by 
counting the number of galaxies within $R_{30}$ that were brighter 
than $r=20$ mag and at least 0.2 mag bluer in $B-V$ than the locus of the 
red galaxy population on the color-projected radius diagram.  The 
results from this calculation are shown Table~3, along with data for 
two Butcher-Oemler clusters, Abell 1942 and Abell 1961, that are 
similar to Abell 2390 in terms of redshift, $R_{30}$, $C_{BO}$, 
richness, and blue fraction\footnote{Abell 1942 was observed with the 
Einstein IPC to have a count rate of 0.015 IPC counts/sec, 
corresponding to a luminosity $\sim 5 \times 10^{44}$\ergps, so its 
X-ray luminosity is certain to be similar to that of Abell 2390.  The 
X-ray luminosity of Abell 1961 is also likely to be similar, based on 
the richness-X-ray luminosity correlation in
\citeN{Ebeling:1993}, which suggests that the cluster has a luminosity 
$>10^{44}$ \ergps.}.  The overall blue fraction in Abell 2390 is not 
only very similar to that in Abell 1961 and Abell 1942, it is also in 
excellent agreement with the canonical $f_b$ versus $z$ relation in 
\citeN{Butcher:1984}.  The {\em gradient} in $f_b$ is also in good 
agreement with that expected from the Butcher
\& Oemler (1984) sample at $z \sim 0.25$, as shown in
Figure~\ref{fig:bocomp}.  However, because of the wide spatial 
coverage of our present dataset only the inner $\sim 25\%$ of the 
Abell 2390 data can be compared to the Butcher \& Oemler sample.  
Nevertheless, the similarity between the composition of the galaxy 
populations in Abell 2390 and typical Butcher-Oemler clusters suggests 
that the mechanisms driving galaxy evolution in these clusters may be 
similar.

\begin{figure} 
\centerline{\psfig{figure=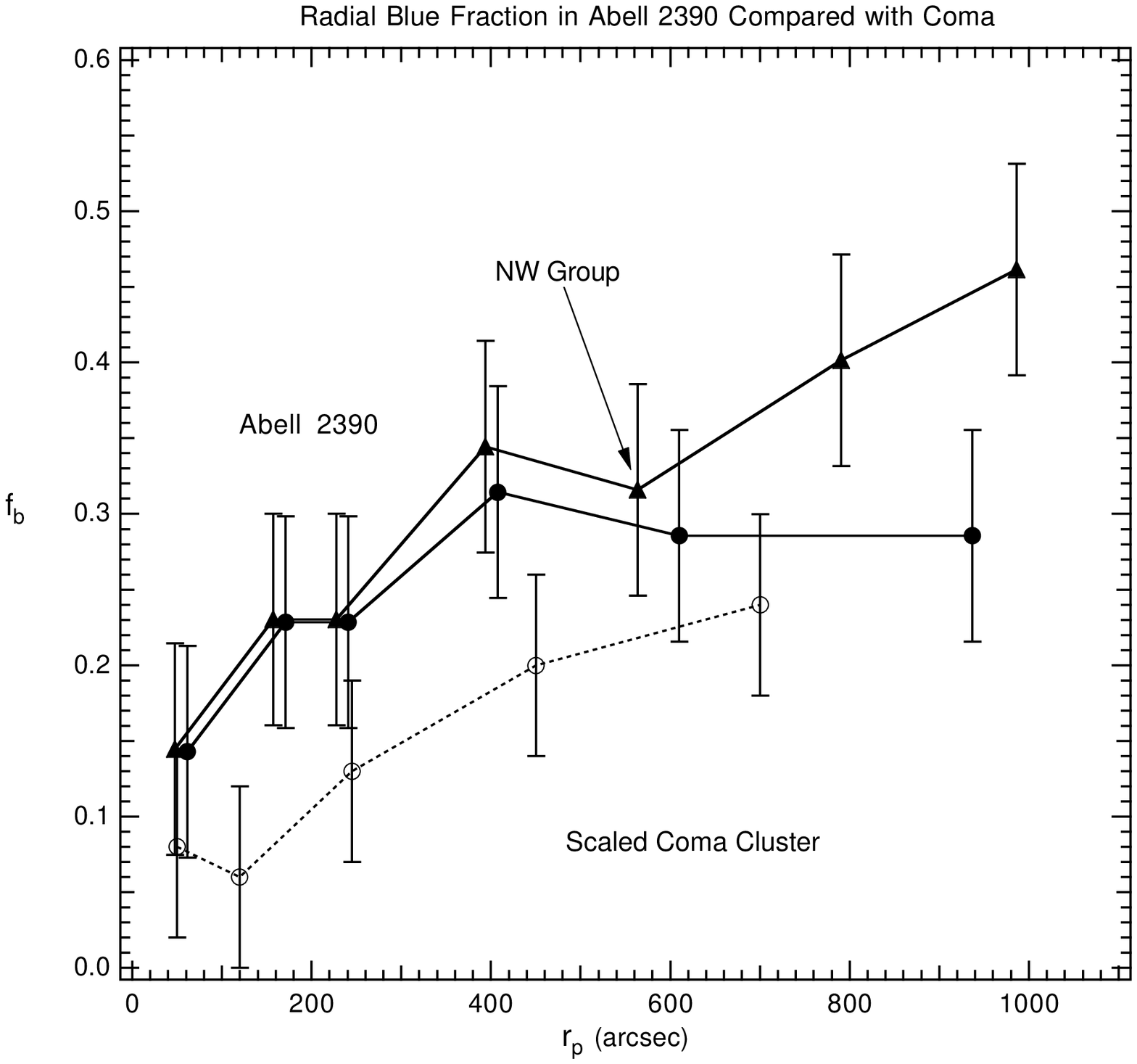,width=5in}}
\caption{ A comparison between the differential blue
    fractions in Abell 2390 and Coma (scaled by a factor of six in
    radius to account for the distance difference between the two
    clusters).  Each bin in both the Coma and Abell 2390 samples
    contains 30-40 galaxies. The Coma data are shown by open circles.
    Abell 2390 points including (filled triangles) and excluding
    (filled points) the ``near-field'' galaxies are shown.}
\label{fig-bluefrac}\end{figure}

In order to compare the blue population at large radii in Abell 2390
to samples other than those of Butcher and Oemler, a comparison was
made between our data and the Coma sample of \citeN{Mazure:1988}.  In
Figure~\ref{fig-bluefrac} we show the differential $f_b$ in Abell 2390
and Coma as a function of projected radius.  In the inner regions of
the clusters contamination by blue galaxies from the periphery of the
clusters may be occuring, although simulations (and the absence of
blue objects near the cluster centers) suggest that such contamination
is small, {\em i.e.} at the level of a few percent.  The differential
blue fraction of the Abell 2390 is 40\% at the periphery of our
dataset, compared with the field value of 58\%, suggesting that even
at large radii red cluster members dominate over blue galaxies.
However, at large radii the blue fraction in Abell 2390 is still a
factor of two larger than that seen in Coma (Figure
~\ref{fig-bluefrac}).  The blue fractions in both samples shown in
Figure ~\ref{fig-bluefrac} were determined in the same manner, with
limiting magnitude cutoffs corresponding to $r=21$ mag at $z=0.23$,
although both curves remain similar even when the limiting magnitude
cutoff is changed by over one magnitude.  The size and coverage of the
\citeN{Mazure:1988} sample is quite comparable with the Abell 2390
data set, both samples having some 200 cluster members over a similar
range of radius in the cluster rest frame.  It is important to note
that the Coma colors show no radial gradient analogous to that seen in
Abell 2390.  However, the slope of the color-magnitude diagram in
Abell 2390 is quite similar to that seen in Coma, where
$\delta(B-V)/\delta(B) = 0.027$ (Caldwell 1995, private communication
based on analysis of data in Godwin \etal [1983]).  Transforming the
slope using a elliptical galaxy spectral energy distribution (see \S5)
gives $\delta(B-V)/\delta(B) \approx 0.019$ for Abell 2390.

\subsection{Cloaked starbursts?}

Since there is so little evidence for starbursting activity in Abell
2390, and since four of the five starburst candidates in the cluster
and near-field appear to be reddish, it is interesting to consider
whether a population of bursting objects may be being cloaked by large
amounts of dust.  The existence of a population of ``cloaked
starbursts'' could alter the conclusions reached in this paper with
regard to the evolutionary history of the cluster population.  It is
therefore significant that the existence of such a population of
``hidden starbursts'' can be ruled out on the basis of deep radio
observations undertaken with the Very Large Array (VLA).  These will
be reported in greater detail elsewhere, and will only be outlined
here.

VLA maps of Abell 2390 were made with the C-configuration at 6cm and
20cm, covering the optically studied area in three and one pointings,
respectively.  The limiting detected flux for sources was $\sim$1 mJy
at 20 cm and 0.3 mJy at 6 cm.  These correspond to a (log) power of
22.4 and 22.0 W/Hz at the cluster redshift, typical of a weak Seyfert
nucleus or strong star-formation.  Thirty two sources were detected at
20 cm and 12 sources at 6 cm, in areas $\sim$3 and 0.5 times that of
the optically covered fields.  Nine of the 12 6 cm sources were
detected at 20 cm.  The central 5 $\times$ 5 arcmin of the cluster
contains nine of the sources, indicating a strong central grouping.
{\em Only two (possibly three) of the cluster members in our
spectroscopic sample were detected in the radio observations, ruling
out the existence of a large population of strongly starbursting
objects hidden by dust.} A summary of the properties of the red
cluster line-emitters suggests that they are a fairly heterogeneous
sample of different types of star-forming galaxies:

{\noindent\bf PPP\# 101084:} This object is the central cD, and is a
strong flat-spectrum point source in the radio with a log power of
24.5 W/Hz.  This object has a peculiar spectrum very unlike that of a
typical elliptical galaxy, but not atypical for cD galaxies in cooling
flow clusters.  It has strong emission [O~II] $ = -111$~\AA\/ as well
as emission at [Ne III] 3867 $ = -4.6$ \AA, weak absorption H$\delta =
0.9$ \AA, emission at H$\gamma = -4.1$ \AA\/ and no detectable G4300
absorption.  The emission line spectrum could arise from gas
photoionized by young stars, a weak AGN, or from gas cooling in a
cooling flow.  Our data do not cover the diagnostic [O~III] or
H$\alpha$ lines, but \citeN{LeBorgne:1991} show this part of the
spectrum, and we conclude that the spectrum is a LINER or H II region
rather than a Seyfert.

{\noindent\bf PPP\# 101033:} This object is a marginally detected
radio source, and has an equivalent width of [O~II] $= -52$ \AA, and
filled-in Balmer line absorption.  Although its projected radius ($r_p
= 78\as$) is small, it has a high velocity relative to the cluster
center, and is likely to be an outer cluster member projected along
the line-of-sight to the core of the cluster.  The optical appearance
of the galaxy is elongated with a low morphological index of $C=0.55$,
suggesting that it is a disk system.  The galaxy appears to be
slightly more extended on one side of the nucleus than the other,
possibly due to a recent merger or interaction.

{\noindent\bf PPP\# 101949:} The spectrum of this object is consistent
with a young population and strong Balmer absorption.  The optical
appearance of this object is very elongated, and the galaxy has a
faint companion.  The morphological index of this object suggests that
it is a late-type system.  This galaxy may be a spiral whose colors
have been reddened by a dust lane.

{\noindent\bf PPP\# 500733:} The galaxy appears isolated, but its
image lies near the edge of the MOS field of view, and is of poor
quality.

\subsection{The role of mergers and interactions}

A major question that has not been addressed is this paper is the the 
role played by tidal events in forming the observed structure and 
galaxy population of the cluster.  Specifically, one would like to 
know if interactions play an important part in defining the cluster 
population (\ie whether or not ellipticals and S0s are forming through 
mergers), and if so how the merger rate is affected by infall into the 
cluster.  Unfortunately a proper investigation of tidal phenomena in 
Abell 2390 requires high resolution imaging data, and our current 
images are not well-suited to this task (due to the comparatively poor 
resolution, coarse sampling, and variable PSF on our CCD frames).  The 
morphological classifier described in this paper provides much 
statistical information on galaxy morphologies in Abell 2390, but the 
nature of the software does not allow the nature of peculiar 
morphology to be quantified.  Future observations and software are 
planned to address this issue.  However, the importance of the 
morphological evidence and the clear peculiarity of some cluster 
members leads us to offer the following qualitative remarks based on 
visual inspection our images.

The visual inspection was made by one of us (JBH), and was based on an
examination of several subsets of the cluster members and near-field
galaxies, selected as follows: (a) strong Balmer absorbers
(H$\delta>4$\AA); (b) strong [O II] emitters ($>$10\AA, 3$\sigma$);
(c) galaxies with high limiting isophotes as determined from our
automated morphological classifier (a sign of overlapping isophotes,
which can be due to crowded fields, peculiar morphology, or image
defects); (d) an inactive sample, with no [O~II] and weak H$\delta$
absorption. Another sample, consisting of (e) all field galaxies
between $0.17<z<0.37$, was also examined as a control.  The following
signs of interaction were looked for: large tidal arms or tails;
double nuclei within an asymmetrical envelope; an asymmetrical light
distribution about the nucleus; warped disks; or bridges to a
companion.  Figure~\ref{fig:mergers} shows a montage of images which
illustrates these features.  The `active' cluster members and field
galaxies - (a) and (b) above - have the same fraction ($\sim$25\%) of
objects showing moderate to strong evidence for interactions, compared
to the $<10{\rm \%}$ of inactive galaxies - (d) above - which show
moderate to strong evidence for interactions. In the field, 18\% of
the sample showed moderate to strong evidence for interactions.
Around 60\% of the `high isophote' galaxies have some morphological
peculiarity suggesting interaction activity, with the rest being
normal but with close companions.  Thus, the automated morphological
classifier does fairly well at picking interaction candidates.
However, most of the high isophote interacting galaxies are not
spectroscopically active.

\begin{figure} \caption{ Montage of R images of the candidate
    merging/interacting galaxies, centered in each box.}
\label{fig:mergers} \end{figure}

The cluster members and near-field galaxies with [O~II] emission were
divided into two groups: the seven lying at least 1$\sigma$ above the
normal-galaxy region in Figure~\ref{fig:o2detns}, and the nine objects
lying within the normal galaxy region (to within error bars).  Of the
high seven, three are probably interacting, two are possibly
interacting, and two look quite normal.  Of the low nine, none are
probably interacting, three are possibly interacting, and six are
normal.  The number of neighbour galaxies (within 10\arcsec) seen in
each group is about the same, averaging about three.  Thus, there is a
good suggestion that the high [O~II] emission may be the result of
star-formation triggered by interactions.

In all, 30 galaxies appear to be good interaction candidates, with 40
more being marginal.  The spatial distribution of these galaxies
roughly follows the density of cluster members, so that they do not
obviously occur either at the outside or the inner parts of the
cluster.  While there are thus definitely {\em some} interacting
galaxies in Abell 2390, the fraction of interacting objects cannot be
accurately determined without higher resolution images.  The cluster
galaxies which are most clearly interacting appear to be red with
strong Balmer absorption and no evidence of [O II] emission.  It is
well-known from the IRAS sample of galaxies that interaction-induced
starbursts are often cloaked with dust, and may not be clearly seen at
optical wavelengths (e.g., \citeNP{Hutchings:1991}).  However, the
radio observations described earlier rule out the existence of a
population of strongly starbursting systems.  Alternatively, these
could be mergers of evolved galaxies that contain only small amounts
of gas that briefly undergo small bursts of star formation, and hence
$\sim 1$ Gyr after the merger their colors are again dominated by the
old stellar component.  Quantitative understanding of the role of
merging in the cluster population clearly must await better data.

\section{CONCLUSIONS}

Our analysis leads to the following conclusions:

1.  Galaxies in the central $0.4$ \hmpc\/ of Abell 2390 are red and
have low dispersions in velocity, color, and spectral line strengths
as well as high central concentrations.  These properties suggest that
they are coeval E/S0 galaxies with ages $\simgt 8$ Gyr (assuming that
their star formation timescales are $\sim 1$ Gyr and that mean
metallicities are approximately solar).  These objects are likely to
be the first generation of galaxies formed in the proto-cluster.

2. Large scale accretion from the field along the West side of the
cluster is suggested by the spatial-velocity structure of the cluster,
and by the the skewed distribution of line emitters (which occur
almost exclusively in the West half of the cluster). The most obvious
structural subgroup in Abell 2390, the NW Group, may be part of this
infall pattern. This group of galaxies is more evolved than its
surroundings and is presumably the core of a smaller cluster being
accreted onto the main component.

3. Radial gradients exist in the colors, spectral features and
morphologies of the cluster galaxies.  These radial gradients are
interpreted as an age sequence in which the mean age of the galaxies
decreases with radius as a consequence of truncated star formation in
spirals accreted from the field.  Many galaxies in the extensive outer
envelope of the cluster have properties intermediate between E/S0s and
field spirals.  These galaxies are analogous to the ``anemic spirals''
seen in local clusters \cite{vandenBergh:1976}, and they are likely to
be transitional objects in an evolutionary sequence in which field
spirals are transformed into into cluster S0s. While the blue fraction
of the cluster rises strongly as a function of radius, even at the
edges of our dataset the galaxy population remains redder than the
field.

4.  Only $\simlt 5$\% of the galaxies in Abell 2390 show signs of star
formation at levels higher than those seen in normal Sbc galaxies.
The large number of H$\delta$ strong objects relative to active
galaxies suggests that star formation has been halted in many
galaxies, and that in most cases this truncation has occurred in
systems that have not undergone starbursts.  This suggests that
truncation in the star formation rates of cluster members is more
closely linked to a gradual mechanism such as stripping by the hot
intracluster medium than to starbursts.  We cannot rule out the
possibility that some ($\sim$25) H$\delta$-strong objects are
post-starburst systems, but if so then the epoch of cluster starbursts
must ended $\simgt 1$Gyr before the epoch of observation.

5. The blue fraction in Abell 2390 is typical of that seen in
``Butcher-Oemler'' clusters at $z\sim 0.25$. The recent HST results of
\citeNP{Couch:1994} and \citeNP{Dressler:1994} suggest that the
increased blue fraction in high-redshift clusters is due to a high
proportion of blue disk galaxies.  If Abell 2390 is really a
``typical'' rich cluster at intermediate redshift, then truncated star
formation leading to the transformation of the blue disks in high
redshift clusters may be the physical mechanism driving the
Butcher-Oemler effect.  Future papers will compare Abell 2390 to other
CNOC clusters (which span the redshift range $0.2 \leq z \leq 0.5$) in
order to determine whether similar mechanisms are driving galaxy
evolution in other X-ray luminous clusters.

\vskip 10pt

\centerline{Acknowledgments}

The data used in this paper forms part of the CNOC (Canadian Network
for Observational Cosmology) study of intermediate redshift clusters.
We are grateful to all the consortium members, and to the CFHT staff
(particularly Telescope Operators Ken Barton, John Hamilton, and
Norman Purves) for their contributions to this project. We also thank
Sidney van den Bergh, Richard Ellis, and Alastair Edge for useful
discussions.

\bibliographystyle{apjv2} \bibliography{astronomy,apjmnemonic}

\end{document}